\documentclass[intlimits,twoside,a4paper]{article}
\usepackage[cp1251]{inputenc}
\usepackage[eqsecnum]{cmpj3}

\issue{2026}{29}{3}{33601}
\doinumber{10.5488/CMP.29.33601}

\title[Phase field modelling of microstructure transformations in Zr--Sn alloy]{Phase field modelling of microstructure transformations in Zr--Sn alloy during irradiation}
\author[V.~O.~Kharchenko, D.~O.~Kharchenko, A.~V.~Dvornichenko]
{ V.~O.~Kharchenko\orcid{0000-0002-0148-6001}\refaddr{label1,label2}\thanks{Corresponding author: \email{vasiliy@ipfcentr.sumy.ua}},
D.~O.~Kharchenko\orcid{0000-0003-3855-313X}\refaddr{label1},
A.~V.~Dvornichenko\orcid{0000-0003-0861-8158}\refaddr{label2}}
\addresses{
\addr{label1} Institute of Applied Physics, National Academy of Sciences of Ukraine, Sumy 40000, Ukraine
\addr{label2} Sumy State University, Sumy 40007, Ukraine
}


\date{Received 15 April 2026; revised 15 June 2026; accepted 16 June 2026; published 28 September 2026}

\begin{document}

\maketitle

\begin{abstract}
In this work we study peculiarities of spatial rearrangement of solute and vacancy concentrations in Zr--Sn alloy at stages of both thermal treatment of solid solution and irradiation of annealed alloy under reactor conditions. To this end, we exploit phase field modelling including reaction rate theory and CALPHAD method combined in one generalized approach. Simulations were done in 3D in order to get a detailed information about rearrangement of solutes and defects in a bulk. We discuss the emergence of the secondary phase at the stage of thermal treatment and its stability under neutron irradiation. We analyze an influence of the irradiation dose onto statistical properties of the secondary phase particles. 
\printkeywords
\end{abstract}

\section{Introduction}

Zirconium-based alloys are widely used as cladding and structural materials in nuclear reactors due to their low neutron absorption cross-section, good corrosion resistance, and satisfactory mechanical properties under extreme operating conditions \cite{Motta2015,Was2007}. 
During thermal treatment and irradiation, zirconium alloys undergo complex microstructural evolution involving phase transformations, solute redistribution, and precipitation of secondary phase (SP) \cite{Griffiths1988,Was2007}. These processes strongly affect macroscopic properties such as strength, corrosion resistance, and dimensional stability \cite{Motta2015}. In particular, precipitation phenomena are governed by the interplay between thermally activated diffusion and irradiation-induced effects, including radiation-enhanced diffusion and radiation-induced segregation  \cite{Was2007,Allen2010}.
Irradiation produces high concentrations of point defects, such as vacancies and self-interstitial atoms, which significantly enhance the atomic mobility and promote non-equilibrium phase transformations \cite{Was2007}. As a result, the kinetics of nucleation, growth, and coarsening of precipitates under irradiation conditions can substantially differ from those observed under purely thermal conditions \cite{Allen2010}.

Tin is one of the principal alloying elements in zirconium alloys and plays a key role in phase stability, diffusion kinetics, determining phase equilibria, microstructural evolution and precipitation~\cite{Motta2015,Barberis2004,Toffolon2000}.
 Experimental studies have shown that Sn affects the solubility limits, defect evolution, and irradiation response in zirconium alloys \cite{Motta2015,Griffiths1988}. In technologically very effective systems such as Zircaloy-2 and Zircaloy-4, tin contributes to the formation and evolution of second phase particles and influences corrosion resistance and irradiation growth \cite{Barberis2004,Toffolon2000}.
Phase diagram studies of Zr--Sn and related multicomponent systems indicate the possibility of forming complex intermetallic phases \cite{Perez2006,Massalski1990}. In particular, metastable phases may arise due to supersaturation, rapid thermal processing, or irradiation-induced effects.
SP formation in Zr--Sn alloys is governed by invariant reactions and temperature-dependent solubility limits. The intermetallic $\eta$-phase (Zr$_5$Sn$_{3+x}$) forms at high temperatures via a eutectic reaction with $\beta$-Zr and  congruently melts at $\sim2260$~K, exhibiting a homogeneity range ($0 \leqslant x \leqslant 1$) \cite{item1}. At lower temperatures, a miscibility gap develops within this phase, leading to separation into compositions close to Zr$_5$Sn$_3$ and Zr$_5$Sn$_4$ \cite{item5,item6}. 

Of particular interest are A15-type intermetallic phases, characterized by a complex cubic structure and typically described by the A$_3$B stoichiometry. While A15 phases are well known in systems such as Nb-Sn, experimental indications suggest that A15-like or structurally related metastable phases may also emerge in zirconium-containing alloys \cite{A15,Massalski1990,Allen2010}.
The A15 phase emerges through a peritectoid reaction between $\beta$-Zr and the $\eta$-phase \cite{item1}. Although initially described as Zr$_4$Sn, it is now associated with a cubic A15 structure exhibiting non-stoichiometry, attributed to Sn vacancies or Zr substitution \cite{item5}. A limited compositional range of this phase has also been reported \cite{item4,item6}. 
At lower temperatures,  a second peritectoid reaction ($\beta$-Zr + A15 $\rightarrow \alpha$-Zr) reduces Sn solubility in $\alpha$-Zr and promotes precipitation of the A15 phase as a SP enriched by tin \cite{item1}. Despite some discrepancies in the reported solubility limits~\cite{item2,item3,item4,item14}, it is generally accepted that a  decreasing temperature enhances SP precipitation.
Precipitation of A15 SP is typically associated with defect accumulation, enhanced diffusion, and local compositional fluctuations.
Despite these indications, the mechanisms governing the nucleation and growth of A15-type phases in Zr--Sn alloys remain poorly understood, particularly under coupled thermal and irradiation conditions. The interplay between radiation-induced defects, solute redistribution, and phase stability presents a complex multiscale problem that requires advanced modelling approaches.

Mathematical modelling has become an essential tool for studying microstructural evolution in irradiated materials. Classical approaches such as rate theory and cluster dynamics have been successfully applied to describe the defect evolution and precipitation kinetics \cite{Was2007,Allen2010}. However, these approaches are limited in their capability to capture a spatially resolved microstructure and complex morphologies.
In this context, the phase-field method emerged as a powerful framework for modelling microstructural evolution. It enables the description of diffuse interfaces, complex morphologies, and the coupling between chemical, elastic, and defect fields \cite{Chen2002,Steinbach2009}. Phase-field models are effectively applied to precipitation and phase transformation phenomena in various material systems \cite{epjb2003,epjb2008}, including zirconium alloys \cite{Bai2017,Kharchenko03082018,kharchenko2023phase,kharchenko2022phase}. 
At the same time, the influence of irradiation onto stability of the complex intermetallic phases, including A15-type structures, its morphology and statistical properties has received limited attention.
Therefore, there is a clear need for the development of a comprehensive phase-field model capable of describing the dynamics of SP precipitation in Zr--Sn alloys under combined thermal treatment and irradiation. Such a model should account for defect-mediated diffusion, non-equilibrium thermodynamics, and the coupling between irradiation damage and phase evolution.

The present work aims at developing a phase-field approach for modelling the nucleation and growth of tin-enriched SP precipitates (such as A15-type phase) in Zr--Sn alloys, with particular emphasis on irradiation-induced effects and non-equilibrium kinetics. We provide 3D modelling of SP precipitation during thermal treatment of the solid solution and analyze microstructure transformations and rearrangement of nonequilibrium defects in Zr--Sn alloy with stable precipitates subjected to a sustained irradiation.

The work is organized in the following manner. In section~\ref{secZrSn-pft} we discuss the phase-field model of microstructure evolution in binary Zr--Sn alloys with point defects. In section~\ref{sec-3} we present results of numerical simulation. We conclude in the last section.

\section{Phase field model of microstructure evolution in  Zr--Sn alloys \label{secZrSn-pft}}

By considering a binary alloy Zr--Sn with atoms of two sorts
we operate with local atomic  concentrations for Zr and Sn as
continuously evolving  fields in space ${\bf r}$ and time $t$ denoted as  $x_{\alpha}(\mathbf{r},t)=\mathcal{N_\alpha}(\mathbf{r},t)/\mathcal{N}$ with $\alpha=\{\rm{Zr, Sn}\}$, where $\mathcal{N}_\alpha(\mathbf{r},t)$ is the number of atoms of the sort $\alpha$, $\mathcal{N}$ is the total number of atoms included in a cubic cell of the size $\ell<\delta$,  where $\delta$ is the interface width. Concentration of point defects $d=\{i,v\}$ ($i,v$ correspond to interstitials and vacancies, respectively) we denote as  $c_{d}(\mathbf{r},t)=\mathcal{N}_d(\mathbf{r},t)/\mathcal{N}$.  The system is considered in a fixed volume  $V$ with periodic boundary conditions.  

\newpage
\subsection{Gibbs energy formalism for Zr--Sn system}

A total Gibbs energy 
\begin{equation}
G_{\rm tot}=G_{\rm Zr-Sn}+\sum_d G_{d}
\end{equation} 
is composed of parts related to the atomic subsystem $G_{\rm Zr-Sn}$ and point defects subsystem $\sum_d G_{d}$. 
The molar Gibbs energy for the atomic subsystem is given by 
\begin{equation}
G_{\rm Zr-Sn}=G_{\rm Zr-Sn}^{\rm ref}+G_{\rm Zr-Sn}^{\rm id}+G_{\rm Zr-Sn}^{\rm ex},
\end{equation}
where $G_{\rm Zr-Sn}^{\rm ref}$ is the reference Gibbs energy, $G_{\rm Zr-Sn}^{\rm id}$ denotes the ideal Gibbs energy
contribution due to random mixing of atoms, $G_{\rm Zr-Sn}^{\rm ex}$ is the excess term defining a deviation from ideality.
The reference Gibbs energy is written in the standard form
\begin{equation}
G_{\rm Zr-Sn}^{\rm ref}=G^{0}_{\rm Zr}x_{\rm Zr}+G^{0}_{\rm Sn}x_{\rm Sn}.
\end{equation} 
Here, we consider  zirconium in $\alpha$-phase $G^{0}_{\rm Zr}=G^{\alpha}_{\rm Zr}$ and assume that  pure tin  $G^{0}_{\rm Sn}=G^{\rm pure}_{\rm Sn}$ dissolved in $\alpha$-zirconium.

The ideal part relates to entropic contribution
\begin{equation}
G_{\rm Zr-Sn}^{\rm id}=RT\left[x_{\rm Zr}\ln x_{\rm Zr}+x_{\rm Sn}\ln x_{\rm Sn}\right],
\end{equation}
where $R$ is the gas constant ant $T$ is the temperature. The excess term has the form 
\begin{equation}
G_{\rm Zr-Sn}^{\rm ex}=x_{\rm Zr}x_{\rm Sn}L_{\rm Sn,Zr},
\end{equation}
where 
\begin{equation}
L_{\rm Sn, Zr}=\sum_n(x_{\rm Sn}-x_{\rm Zr})^nL^n_{\rm Sn,Zr}(T).
\end{equation}
The coefficients $L^n_{\rm Sn,Zr}$ are obtained from least-squares fitting of
experimental data. They may be temperature dependent: $
L^n_{\rm Sn,Zr}(T)=A+BT$.
Mainly regular interaction parameter relates to  $n = 0$, and
sub-regular interaction corresponds to $n = 1$. They are used to define  the excess Gibbs energy as:
\begin{equation}
G_{\rm Zr-Sn}^{\rm ex}=x_{\rm Zr}x_{\rm Sn}\left[L^0_{\rm Sn,Zr}+(x_{\rm Sn}-x_{\rm Zr})L^1_{\rm Sn,Zr}\right].
\end{equation}
For the defect subsystem, the contribution $G_d$ for the $d$-type defect has the following form:
\begin{equation}
G_{d}=G^f_{d}+G_{d}^{\rm id}+G^{\rm int}_{d}.
\end{equation}
Without loss of generality, one can define the formation energy of defects $G_d^f$ through  its formation energies $G_d^{f,\rm Zr}$ in pure Zr  and  $G_d^{f,\rm Sn}$ in pure tin, and the nominal concentration of zirconium $\overline{x}_{\rm Zr}$ and tin $\overline{x}_{\rm Sn}$ in the form
\begin{equation}\label{EqGv_ZrSn1}
G_d^f=G_d^{f,\rm Zr}\overline{x}_{\rm Zr}+G_d^{f,\rm Sn}\overline{x}_{\rm Sn}.
\end{equation}
The entropic contribution is   $G_d^{\rm id}=RTc_d\ln c_d$. 
Interactions of point defects with zirconium and tin  atoms can be described by the Gibbs component: 
\begin{equation}
\begin{split}
G^{\rm int}_{d}=c_d\left[x_{\rm Zr}G_{d-Zr}^{\rm int}+x_{\rm Sn}G^{\rm int}_{d-Sn}\right],
\end{split}
\end{equation}
where for the defect-atom interaction energies we use:
\begin{equation}
G^{\rm int}_{d-Zr}=\frac{G_{\rm coh}^{\rm Zr}+G_d^{f,\rm Zr}}{Z},\quad    
G^{\rm int}_{d-Sn}=\frac{G_{\rm coh}^{\rm Sn}+G_d^{f,\rm Sn}}{Z}.
\end{equation} 
Here, $G_{\rm coh}^{\rm Zr,Sn}$ denote the corresponding cohesive energies, $Z$ relates to the coordination number. 

The corresponding Gibbs energy functional
\begin{equation}
\mathcal{G}=\frac{1}{V_m}\int{\rm d}V\left[ G_{\rm tot}(\{x_{\alpha}\}, \{c_d\})+\sum_\alpha\kappa_{\alpha}(\nabla c_{\alpha})^2+\sum_d\kappa_d(\nabla c_d)^2\right] 
\label{Gibbs}
\end{equation}
includes gradient energy terms $\kappa_\alpha$, $\kappa_d$. In further consideration we put $\kappa_\alpha=L^0_{\rm Sn,Zr} \ell^2/6$,  
$\kappa_d=\kappa_\alpha/4$;  $V_m$ is the molar volume.

In multicomponent models, the mass conservation law $\sum_\alpha x_\alpha=1$ can be used to eliminate one concentration field.
In the considered system, one can express  $x_{\rm Zr}$ through  $x_{\rm Sn}$ as: $x_{\rm Zr}=1-x$, where $x\equiv x_{\rm Sn}$. 
By taking into account that tin atoms are capable of trapping vacancies due to the attractive tin-vacancy interaction (see reference~\cite{WU2021101765}) and assuming that interstitials with concentration $c_i\ll c_v$ are homogeneously distributed in a bulk, one can consider equilibrium vacancies as point defects only. 
In such a case, the dynamics of the complete  system in the equilibrium conditions is governed by equations
\begin{equation}\label{prepZrSn}
\begin{split}
&\partial_t x=\nabla\cdot M_x\nabla\frac{\delta \mathcal{G}}{\delta x},\\
&\partial_t c_v=\nabla\cdot L_v\nabla\frac{\delta \mathcal{G}}{\delta c_v},
\end{split}
\end{equation}
where $M_x$ and $L_v$ are the mobilities of the atoms and vacancies, respectively. Here, we use the standard relations $M_x=x(1-x)\left[
xM_{\rm Zr}+(1-x)M_{\rm Sn}\right]$, where $M_\alpha=D_\alpha^0/RT$ is defined through the diffusivities $D_\alpha^0$ of $\alpha$-atoms, and 
and  $L_v=c_vD_v/RT$, where  $D_v$ is the diffusivity of vacancies. In further consideration, we assume $D_{\rm Zr}^0=D_vc_{v0}$, where $c_{v0}$ is the equilibrium concentration of vacancies. The system (\ref{prepZrSn}) with (\ref{Gibbs}) can be used to modelize microstructure transformations in Zr--Sn alloys with vacancies at equilibrium conditions (thermal treatment).

\subsection{Irradiation influence onto microstructure evolution in  Zr--Sn alloys}

By considering the system under irradiation, one takes into account the ballistic mixing  and   radiation enhanced diffusion of atoms. In this case, the evolution equation for the tin concentration (first equation in equation~(\ref{prepZrSn})) is generalized by adding the term  responsible for ballistic mixing, describing an interchange of atoms with a frequency $\Gamma$ on the average atomic relocation distance $\varrho$ \cite{martin1984phase}: \mbox{$\Gamma(\langle x\rangle_\varrho-x)$}. Here, the average $\langle x\rangle_\varrho$ is taken over a distribution $w_\varrho(\mathbf{r})$ of ballistic exchange distances under irradiation~\cite{enrique2000compositional,enrique2001compositional,
enrique2003simulations,enrique2004nonequilibrium,
demange2017prediction}:
$\langle x\rangle_\varrho=\int w_\varrho(\mathbf{r}-\mathbf{r}')x(\mathbf{r}'){\rm d}\mathbf{r'}$. In our modelling, we use the typical model $w_\varrho(r)=({1}/{2\piup \varrho^2})\re^{-r/\varrho}$~\cite{kharchenko2022phase,kharchenko2024modeling}.
The atomic jump frequency $\Gamma=\mathcal{K}A_{\rm irr}$ is defined through the damage rate $\mathcal{K}$, computed according to the NRT standard \cite{norgett1975proposed},  and a constant $A_{\rm irr}\approx 50$ for neutron irradiation~\cite{ke2019flux}.

At irradiation the nonequilibrium point defects (both vacancies and interstitials) can be produced, move to sinks and can recombine \cite{Was}. 
In such a case, the evolution equation for concentration of each type of point defects becomes as follows 
\begin{equation}
\partial_t c_{d}=\mathcal{R}_d-\nabla\cdot \mathbf{J}_d,
\end{equation}
where $\mathcal{R}_d$ is the reaction term and $\mathbf{J}_d$ relates to the diffusion flux of point defects of $d$-type.  
The reaction part has the form $\mathcal{R}_d=\mathcal{K}(1-\varepsilon_d)-D_dk^2c_d-\alpha_r c_ic_v$. Here,  $\varepsilon_d$ is the efficiency of $d$-type defect clustering, $D_d$ is the point defect diffusivity. Here, we consider the network dislocations and both vacancy and interstitial loops as major sinks of point defects. In order to simplify our description,  we assume that network dislocation density remains constant during irradiation, whereas vacancy and interstitial loop densities evolve. The second term describes the absorption of point defects  by sinks with strengths $k^2=Z_N\rho_N+Z_I{\rho_{I}}+Z_V{\rho_{V}}$, where  $\rho_N$ is the network dislocation density, $\rho_{I,V}$ relate to  interstitial and vacancy loop density.  Bias factors defining the preference of the absorption of defects by sinks are: $Z_N=1.0$, $Z_V=1.1$, $Z_I=1.2$ \cite{Was}. 
The third term  corresponds to recombination of point defects with  the rate $\alpha_r=4\piup r_0D_i/\Omega_0$ expressed through the defect recombination radius  $r_0\simeq (3-5)a$ ($a$ is the lattice constant) and the atomic volume $\Omega_0$. 

In order to simplify the mathematical model, the following assumptions can be made. From the  dynamical equilibrium condition  $D_ic_i\simeq D_vc_v$, one can relate concentration of interstitials to concentration of vacancies \cite{Turkin2006} by taking into account  fast migration of interstitials to  sinks \cite{golubov2012radiation} compared to vacancies. The relation $D_i\gg D_v$ yields $\nabla\cdot \mathbf{J}_i \approx 0$.  In this case, one can focus only on the local rearrangement of vacancies \cite{kharchenko2021stability,Kharchenko03082018} with vacancy diffusion flux  $\mathbf{J}_v$.  In general, the diffusion flux $\mathbf{J}_v$  is composed of the  thermodynamics  related to the corresponding Gibbs energy \emph{via} standard Onsager relation and the part relevant to interaction of vacancies due to elastic deformation of the lattice \cite{Kharchenko03082018,CondMat2013}. In such a case, one gets: $\mathbf{J}_v=\mathbf{J}_v^{\rm th}+\mathbf{J}^{\rm int}_v$, where $\mathbf{J}_v^{\rm th}=-L_v\nabla (\delta \mathcal{G}/\delta c_v)$ is governed by thermodynamics and  
$$\mathbf{J}^{\rm int}_v=L_v \frac{EV_m}{3(1-2\nu)}\nabla \left[1+\frac{r_{v0}^2}{3}\frac{1+\nu}{1-\nu}\nabla^2\right]c_v$$
 takes into account the  interactions of vacancies  through elastic deformation of the lattice   \cite{mirzoev1996laser,kharchenko2016patterning,
kharchenko2016modeling,kharchenko2020phase,
kharchenko2021stability,CondMat2013}.  Here, $E$ is the elastic modulus, $\nu$ is the Poisson ratio, $r_{v0}\approx (2$~{to}~$3)a$ denotes the mean square distance
between the produced defect and the atom of the matrix. 

By taking into account radiation-enhanced diffusion, one can put  $D_{\alpha}=D_{\alpha}^0+D_v \langle c_v\rangle$, where $D^0_{\alpha}$ relates to atomic diffusivity under equilibrium conditions, $\langle c_{v}\rangle$ denotes mean concentration of nonequilibrium  vacancies.
The spatio-temporal evolution of atomic and defect subsystems is described by the following coupled equations
\begin{equation}
\begin{split}
\partial_t x&=\nabla\cdot M_x\nabla\frac{\delta \mathcal{G}}{\delta x}+\Gamma(\langle x\rangle_{\varrho}-x),\\
\partial_t c_v &=\nabla\cdot L_v\bigg[\nabla\frac{\delta \mathcal{G}}{\delta c_v} - \frac{EV_m}{3(1-2\nu)}\nabla \bigg(1+\frac{r_{v0}^2}{3}\frac{1+\nu}{1-\nu}\nabla^2\bigg)c_v\bigg]+ \mathcal{K}_{v}(1-\varepsilon_v)-D_vk^2_v c_v-\frac{\alpha_r D_v}{D_i} c_v^2.
  \end{split}
\label{irrZrSn}  
\end{equation}
We describe dynamics of interstitial and vacancy loop densities $\rho_{I,V}$  by using reaction rate
theory with the model proposed in reference~\cite{89REDS2020}, by taking constant values of the loop number
densities for interstitial and vacancy loops $N$. In
our case, the governing equations for loop densities take the form \cite{51REDS2020,89REDS2020,kharchenko2020phase}:
\begin{equation}
\begin{split}
\partial_t\rho_I&=\frac{1}{r_{0I}|{\bf b}|}\left(
\mathcal{K}\varepsilon_i-Z_I\rho_I\left[D_v(\langle c_v\rangle-c_{iL})-D_i\langle c_i\rangle\right]
\right),\\
\partial_t\rho_V&=\frac{1}{r_{0V}|{\bf b}|}
\left(
\mathcal{K}\varepsilon_v-Z_V\rho_V
\left[D_i\langle c_i\rangle-D_v(\langle c_v\rangle-c_{vL})\right]
\right).
\end{split}
\label{loops}
\end{equation}
Here, $|{\bf b}|$ is the Burgers vector modulus, $r_{0d}$ ($\{d=i,v\}$) denotes the initial interstitial/vacancy loop radius; $c_{dL}$ denotes the equilibrium vacancy concentration near the vacancy and interstitial dislocation loops, respectively, $\langle c_i\rangle=(D_v/D_i)\langle c_v\rangle$ is the mean interstitial concentrations. We take into account that the vacancy loops start to grow after the dose 3~dpa \cite{rrtloops,wu2020dislocation}. Equations (\ref{irrZrSn}) and (\ref{loops}) can be used to modelize spatio-temporal evolution of the annealed Zr--Sn alloy under irradiation. 

It should be noted that at extremely high irradiation doses and temperatures,  interdiffusion processes can certainly be important in the kinetics of nonequilibrium processes. In the present model, interdiffusion is not introduced as an additional phenomenological flux term, but is inherently accounted for through the variational thermodynamic structure of the model and the corresponding mobility by taking into account both radiation-induced ballistic mixing and radiation-enhanced diffusion via vacancy dynamics. Therefore, the combined thermodynamic and irradiation-driven mechanisms capture the essential physics of interdiffusion in the system.

\section{Results and discussions}
\label{sec-3}

In this work we consider the alloy Zr--10\%Sn by fixing the nominal concentration of tin in Zr--Sn alloy as $\overline{x}_{\rm Sn}=0.1$.  
We start with simulations of the phase decomposition in Zr--10\%Sn model system during thermal treatment of solid solution at fixed temperature. Finally, we discuss the influence of the sustained  irradiation onto microstructure of annealed Zr--10\%Sn alloy at a fixed damage rate. In order to perform numerical simulations, it is more convenient to  move to dimensionless quantities: $\mathbf{r}'=\mathbf{r}/\ell$ and $t'=t/\tau$, where $\tau=\ell^2/D_{v}$. Modelling procedure is provided on the square lattice $N\times N\times N$ of the size $N=128\Delta x$ ($\Delta x=1$) with the dimensionless time step  0.001 for insuring
the accuracy in the solving procedure. Boundary conditions are periodic. 
Initial conditions are defined according to a specific problem of modelling. For thermal treatment of the solid solution we use the random distribution of the concentration field $x({\bf r})$ around nominal value  $\overline{x}_{\rm Sn}$. At the stage of modelling irradiation influence,   the initial conditions are taken as a  microstructure related to the final stage of thermal treatment modelling.  This  choice of initial conditions for irradiation influence modelling allows one to provide a detailed analysis of stability of the pre-irradiated microstructure of the alloy during irradiation dose accumulation. The semi-implicit
Fourier spectral method \cite{Chen98,BinerBook} was adopted to solve the system of dynamical equations (\ref{prepZrSn}) at the stage of thermal treatment and (\ref{irrZrSn}) at the stage of irradiation. All parameters used in simulations are collected in table~\ref{tab01}.

\begin{table*}[h]
	\caption{Material parameters used in simulations.}
	 \label{tab01}
	\centering
	\begin{small}
		\begin{tabular}{c|c|c|c}
			\hline\hline
			Parameter & Value & Dimension& Reference\\
			\hline\hline
			Lattice parameters for Zr ($a$, $c$)& $(3.2, 5.14)\cdot 10^{-8}$& cm&\\
			Atomic volume ($\Omega_0$)& $3.32\cdot 10^{-23}$& cm$^{3}$&\\
			Equilibrium vacancy concentration ($c_{v0}$) & $0.54\re^{-E_v^{f,\rm Zr}/k_{\rm B}T}$& at. fract.& \cite{Turkin2006}\\
			Vacancy formation energy  ($E_v^{f,\rm Zr}$)&$1.8$& eV&\cite{Turkin2006}\\
			Vacancy formation energy  ($E_v^{f,\rm Sn}$)&$0.94$ & eV&\cite{AI1}\\
			Cohesive energy ($E_{\rm coh}^{\rm Zr}$)& 6.3& eV & \cite{KPN1995}\\
			Cohesive energy ($E_{\rm coh}^{\rm Sn}$)& 3.15& eV & \cite{EcohSn}\\
			\hline\hline
			Reference Gibbs energy ($G^{\alpha}_{\rm Zr}$)&$-7829+125.649 T$&&\\&$-24.1618 T\ln T-0.00437791 T^2$&J/mol&\cite{Calphad_potentials}\\
			Reference Gibbs energy ($G^{\rm pure}_{\rm Sn}$) &
			$8555$&J/mol&\cite{Calphad_potentials}\\
			Interaction coefficient ($L^{0}_{\rm Sn,Zr} $)&$-79 959.713 + 20.00604 T$&J/mol&\cite{A15}\\
			Interaction coefficient ($L^{1}_{\rm Sn,Zr} $)&$-99 146.4844$&J/mol&\cite{A15}\\
			Spatial scale ($\ell$)         & 6.0    & nm& This work\\
			Sn diffusion coefficient in $\alpha$-Zr ($D^0_{\rm Sn}$) & $3.12\times 10^{-6}\re^{-2.74\text{ eV}/k_{\rm B}T}$&m$^{2}$/s&\cite{SnDiff}\\
			Diffusivity of vacancies ($D_v$)  & $2.2\cdot 10^{-2} \re^{-0.93 \text{ eV}/k_{\rm B}T}$&  cm$^{2}$/s & \cite{CB2005}\\
			Diffusivity of interstitials ($D_i$)  & $3.5\cdot 10^{-4} \re^{-0.06 \text{ eV}/k_{\rm B}T}$&  cm$^{2}$/s & \cite{CB2005}\\
			\hline\hline
			Clustering efficiencies ($\varepsilon_{i}$, $\varepsilon_{v}$)    & 0.67, 0.7& &\cite{kharchenko2021stability,kharchenko2022phase} \\
			Dislocation network density ($\rho_N$)&$10^{10}$&cm$^{-2}$&\cite{Was}\\
			\hline\hline
		\end{tabular}
	\end{small}
\end{table*}

\subsection{Thermal treatment of solid solution}

\begin{figure}[h]
\centering
\includegraphics[scale=0.8]{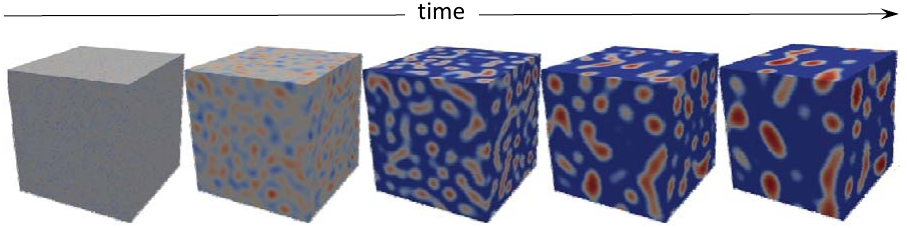}
\caption{(Colour online) Snapshots of the evolution of the system Zr--10\%Sn starting from the quasi-homogeneous solid solution.}
\label{prep1}
\end{figure}

In order to perform numerical simulations of the phase separation process in the studied system Zr--Sn with the nominal concentration $\overline{x}_{\rm Sn}$ during thermal treatment of solid solution, we numerically solve equations (\ref{prepZrSn}) at a fixed temperature $T=550$~K. For initial conditions, we use: $\langle x(\mathbf{r},t=0)\rangle=\overline{x}_{\rm Sn}$, $\langle c_{v}(\mathbf{r},t=0)\rangle=c_{v0}$, 
$\langle (\delta x(\mathbf{r},t=0))^2\rangle=0.01\overline{x}_{\rm Sn}$,  
$\langle (\delta c_{v}(\mathbf{r},t=0))^2\rangle=0.1{c}_{v0}$.  Typical evolution scenario of the thermal treatment of the solid solution is shown in figure~\ref{prep1}. It is seen that starting from the totally homogeneous configuration, new phase (domains related to A15-phase) emerges where concentration of tin is larger than in the bulk. These domains grow in time by absorbing tin from the bulk.  

\begin{figure}[h]
	\centering
	\includegraphics[width=0.5\textwidth]{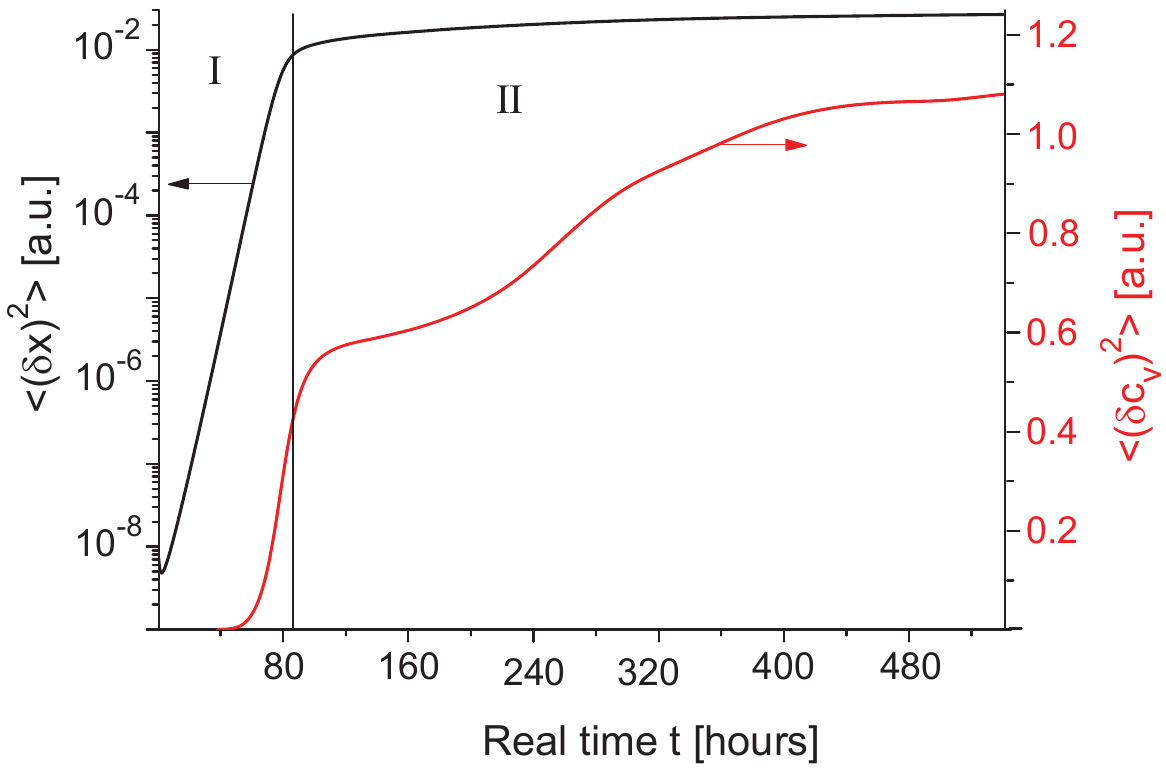}
	\caption{(Colour online) Dynamics of dispersions of Sn concentration and equilibrium vacancy concentration for  Zr--10\%Sn alloy annealed at $T=550$~K.}
	\label{disp}
\end{figure}

The detailed information about the dynamics of the alloy component  rearrangement can be studied by considering the dispersions of concentration  of tin and vacancies: $\langle(\delta x)^2 \rangle=\langle(\overline{x}_{\rm Sn}-x)^2\rangle$ and  $\langle(\delta c_{v})^2 \rangle=\langle(c_{v0}-c_{v})^2\rangle$ shown in figure~\ref{disp}. Here, dispersions of the tin  and vacancies are shown in the left-hand and right-hand axis, respectively. The total  computation time corresponds to 566 hours of real time  annealing according to the  scaling used.
The growing time dependence of the dispersion of the corresponding concentration  field means a spatial ordering of this  field.  
In figure~\ref{disp} one can issue two main stages: (I) a growth stage and (II) a coarsening stage. The growth stage relates to the time interval when both 
dispersions grow (around 80 hours of real time of annealing). Here, the dispersion of tin concentration grows rapidly, meaning a fast rearrangement of atoms due to their interactions and diffusion. An increase in the dispersion of 
equilibrium vacancies field means that the vacancies which were initially homogeneously distributed in a bulk, nucleate by forming areas of dense phase enriched by vacancies.  It means that rearrangement of tin concentration is strongly related to the rearrangement of vacancies.  At this stage, a new phase emerges by  absorbing atoms from the bulk, and the domains of this phase grow in size. At  the coarsening stage, the growth speed of both $\langle(\delta x)^2 \rangle$ and $\langle(\delta c_v)^2 \rangle$ decreases. Here, the   dynamics of the system is described by interaction of domains of new phases. At the long-time limit (over 550 hours of annealing), one has the quasi-stationary 
regime, when both dispersions do not change. This means that all 
ordering processes have finished and the further annealing will not change the spatial 
configuration of the  alloy studied. 

\begin{figure}[h]
	\centering
	\includegraphics[width=0.5\textwidth]{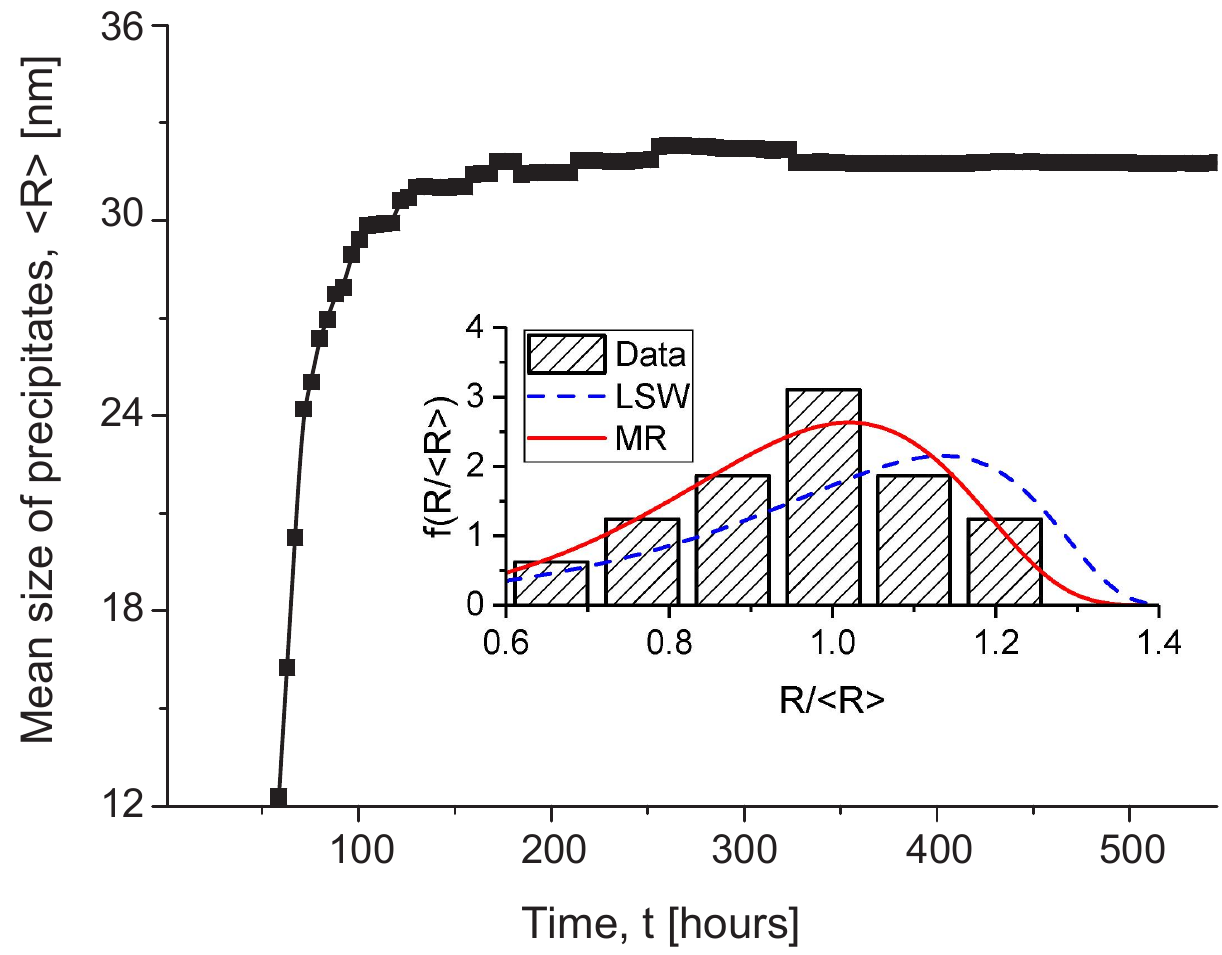}
	\caption{(Colour online) Dynamics of mean size of precipitate of Sn-enriched SP for  Zr--10\%Sn alloy during thermal treatment at $T=550$~K. Distribution of SP particles over sizes after 560~hours annealing is shown in the inset.}
	\label{prep3}
\end{figure}

In order to analyze the dynamics of the mean size of SP particles enriched by tin concentration, we proceed in the following manner. By taking into account that A15 phase is associated with the configuration Zr--$x$Sn where $x\in[1/5,3/8]$ \cite{item5,item6,A15}, we use the mean threshold $x^{th}=0.3$ in order to separate SP particles with $x({\bf r})>x^{th}$ from the matrix phase with $x({\bf r})<x^{th}$. Next here and thereafter by exploiting the machine learning method, known as
density-based spatial clustering of applications with noise
(DBSCAN)  \cite{schubert2017dbscan} with periodic boundary conditions, we calculate the volume $V_i$ of each SP particle. Finally, by using the standard relation $V_i=4\piup R_i^3/3$ we calculate the mean size $R_i$ (radius of the sphere of the equivalent volume) of each particle of SP and find the mean linear size of particles as $\langle R\rangle=(1/n)\sum_iR_i$ during simulations, where $n$ is the total number of particles. 
The dynamics of mean size of particles of the SP is shown in figure~\ref{prep3}. 
It is seen that during the system evolution, the particle size of SP enriched by tin concentration grows fast at early stage when spatial instability in the system is essential. At late stages, its dynamics is slowed down. Here, slow coalescence regime occurs when particles interact, and the growth of particles here is of diffusion character only. After 300 hours, the system attains a quasi-stationary regime and is characterized by a stable microstructure. The corresponding distribution $f(R/\langle R\rangle)$  of SP particles over the sizes is shown in the inset in figure~\ref{prep3} after long-term annealing. It follows that this distribution is slightly skewed with the longest tail at $R<\langle R\rangle$, but remains of symmetric form and   most of the  particles are characterized by the size around the mean value $\left<R\right>\simeq32$~nm with small dispersion. It follows that the 
Lifshitz--Slyozov--Wagner distribution in its original form \cite{LSW1,LSW2} does
not fit directly the numerical data (see dash curve in the inset in figure~\ref{prep3}). A more accurate fitting (see solid curve in the inset in figure~\ref{prep3}) is given by using Marqusee and Rose approach \cite{MR}.
The volume fraction of SP $V_{fr}$ for the used threshold $x^{th}$ calculated as a ratio between the whole volume of SP and the volume of the computational system gives $V_{fr}=16\%$ which is close to the result that can be obtained by analysing the experimental microstructure shown in reference~\cite{A15}. The obtained value of the mean size at low temperature ($T=550$~K) is less than the experimental observations provide (see reference~\cite{A15}) at a high temperature. This difference looks natural because the mean size of precipitates increases with the increasing annealing temperature, as shown by phase field modelling for Zr-Nb-Sn alloys \cite{mifint2025} and Fe-Cr-Al alloys \cite{frontiers}.

\begin{figure}[h]
	\begin{center}
		a)\includegraphics[width=0.3\textwidth]{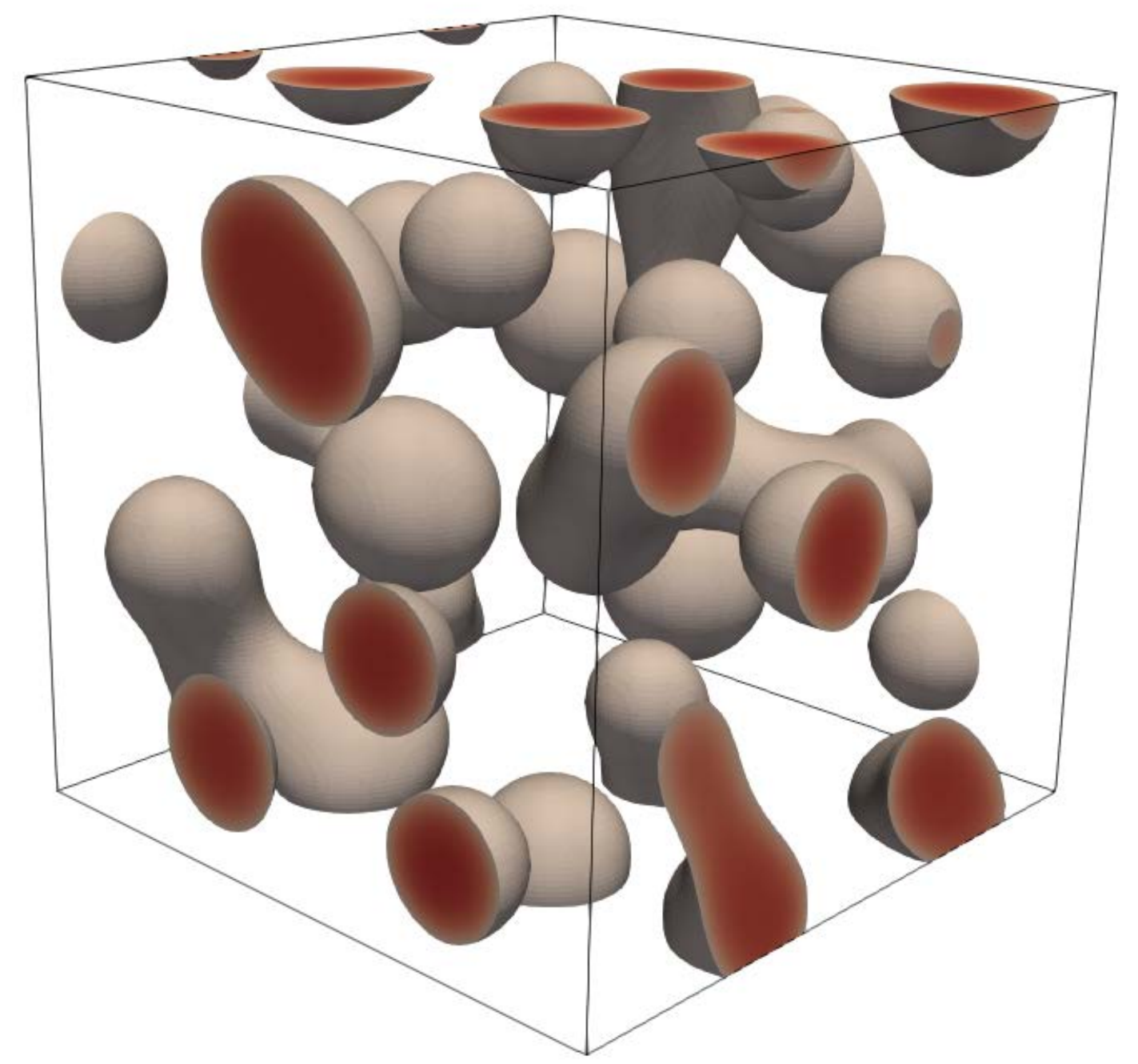}
		b)\includegraphics[width=0.3\textwidth]{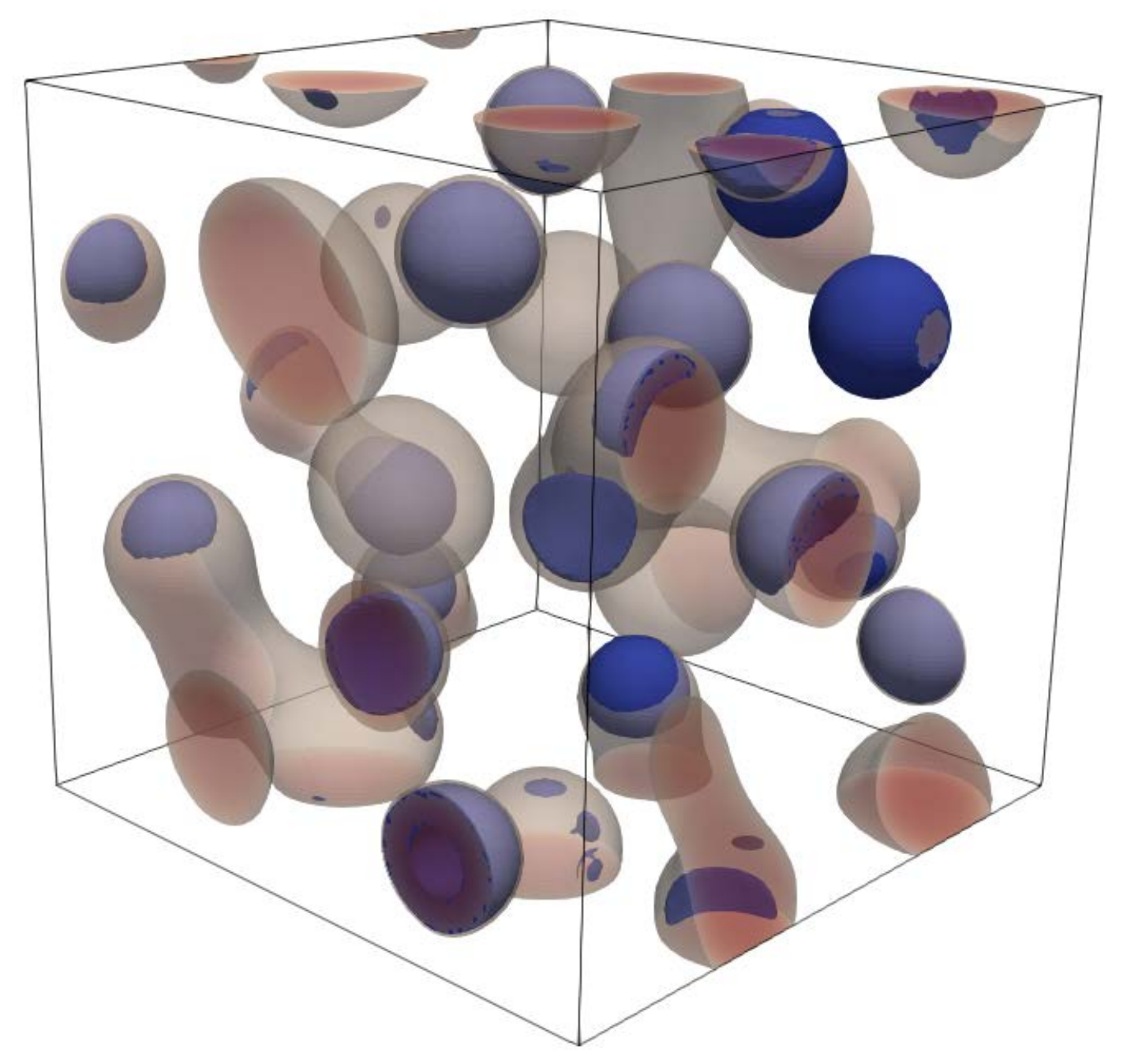}
	\end{center}
	\caption{(Colour online) Typical snapshots of the annealed alloy Zr--Sn: a) domains of SP, enriched by Sn atoms; b) distribution of 
		concentration of equilibrium vacancies in the vicinity of SP domains.}\label{prep2}
\end{figure}

The spatial configuration of SP particles after a long-term annealing, that corresponds to the $f(R/\langle R\rangle)$ in the inset in figure~\ref{prep3}, is shown in figure~\ref{prep2}a. One sees that the domains of SP are of different shape and of different size. In figure~\ref{prep2}b we show the combined configuration 
of two fields: concentration of tin with $x({\bf r})>x^{th}$ is shown with small opacity in red color, 
concentration of equilibrium vacancies is shown by blue color. It follows that the 
equilibrium vacancies segregate mostly 
on the interfaces with large curvature and inside the domains of SP. This is related to attractive Sn-vacancy interaction resulting in trapping the vacancies by Sn atoms \cite{AI1}. The configuration shown in figure~\ref{prep2} will be used as a ``target'' to perform numerical simulations 
of the microstructure transformations and spatial rearrangement of the concentration of the alloying elements and non-equilibrium vacancies induced by the neutron irradiation influence.

\subsection{Irradiation of the annealed alloy}

In order to modelize the microstructure transformations in the annealed alloy during the  irradiation action, we numerically solve  the system (\ref{irrZrSn}) together with the  equations for loops densities (\ref{loops}). Irradiation of the annealed alloy under reactor conditions was simulated at a dose rate $\mathcal{K}=10^{-6}$~dpa/s and temperature  $T=550$~K. The initial conditions for both fields $x({\bf r})$ and $c_v({\bf r})$ are taken from figure~\ref{prep2}b. 

\subsubsection{Spatial rearrangement of the atomic subsystem} 

\begin{figure}[h]
\centering
a)\includegraphics[width=0.33\textwidth]{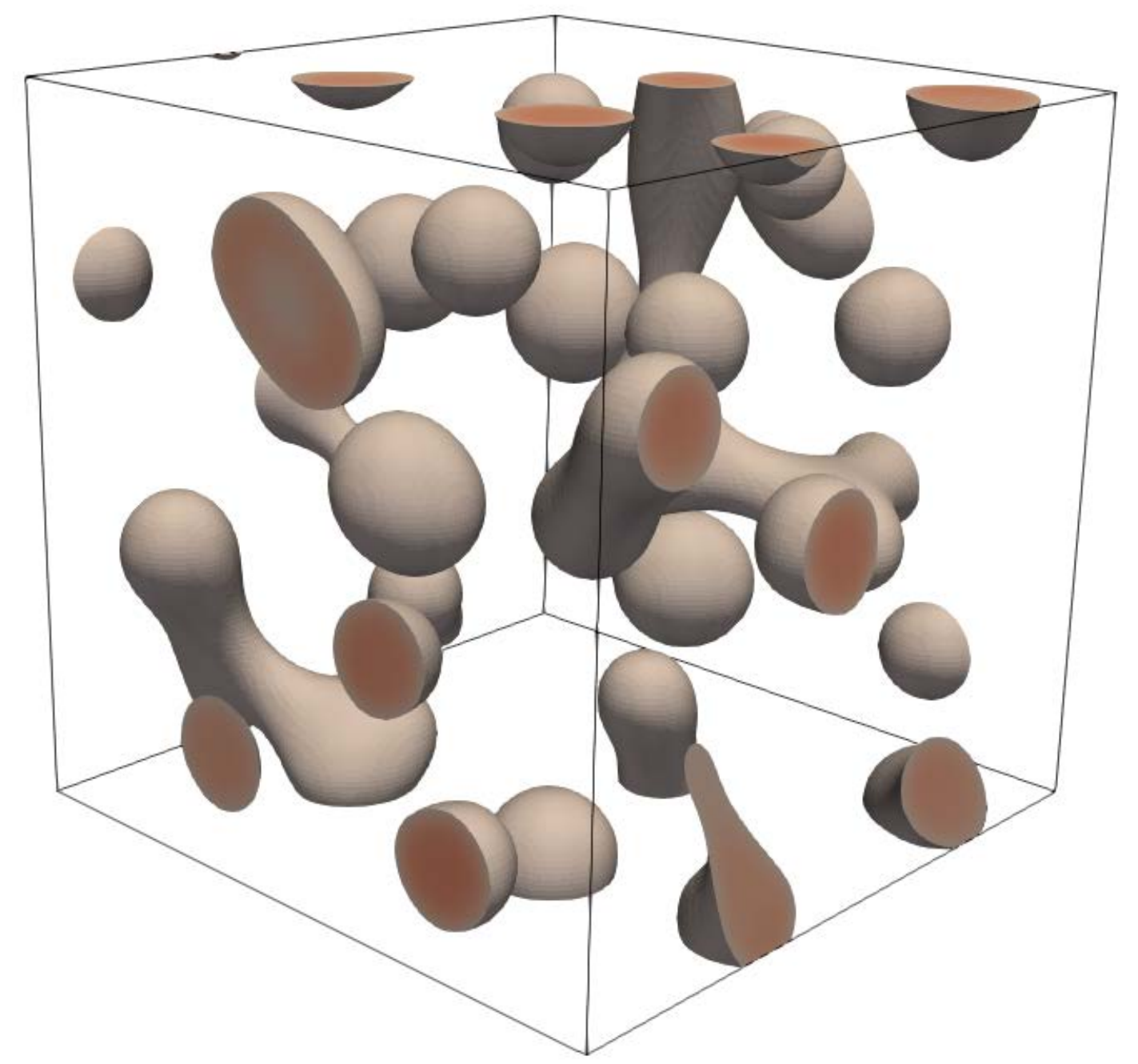}
b)\includegraphics[width=0.33\textwidth]{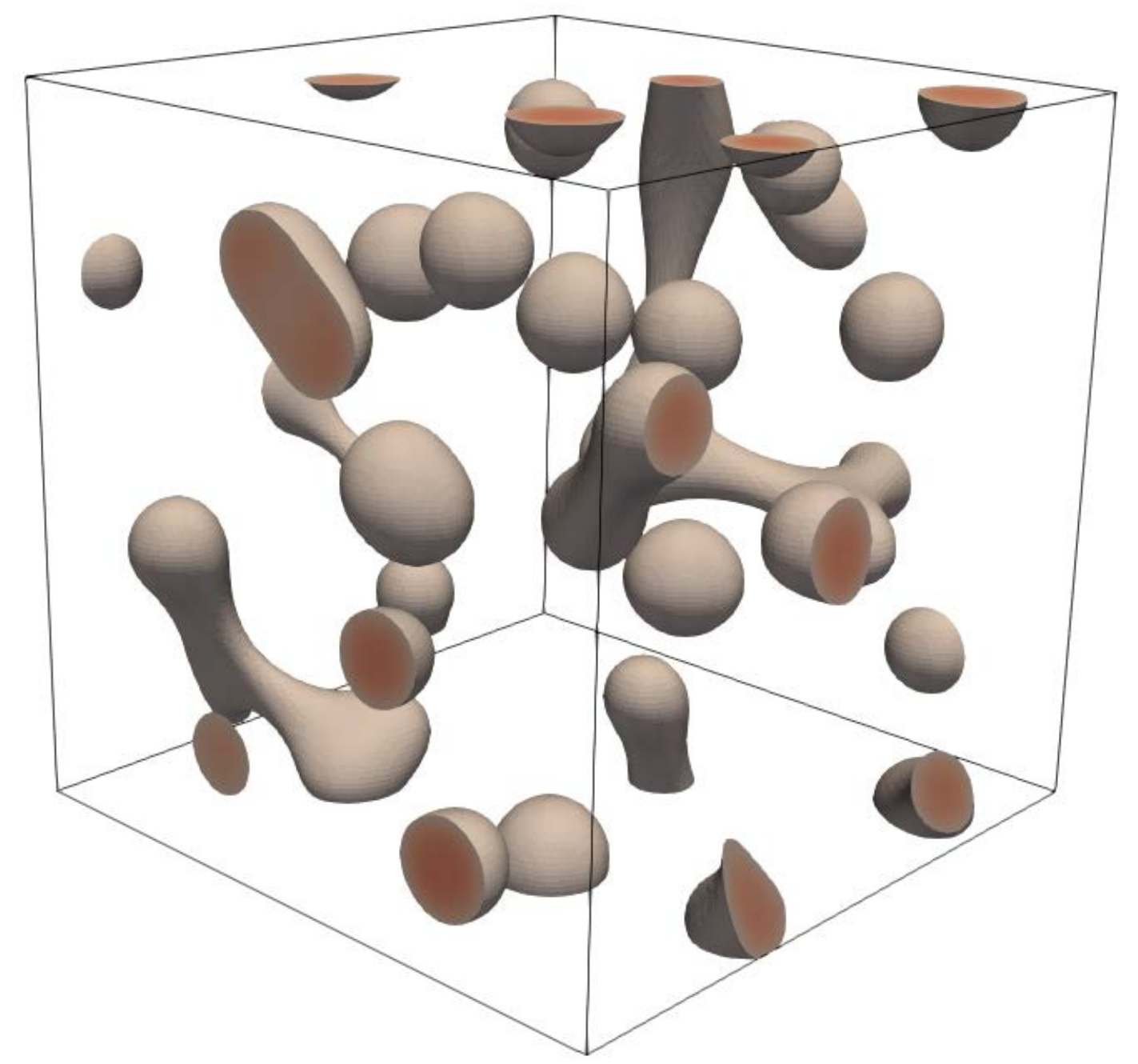}\\
c)\includegraphics[width=0.33\textwidth]{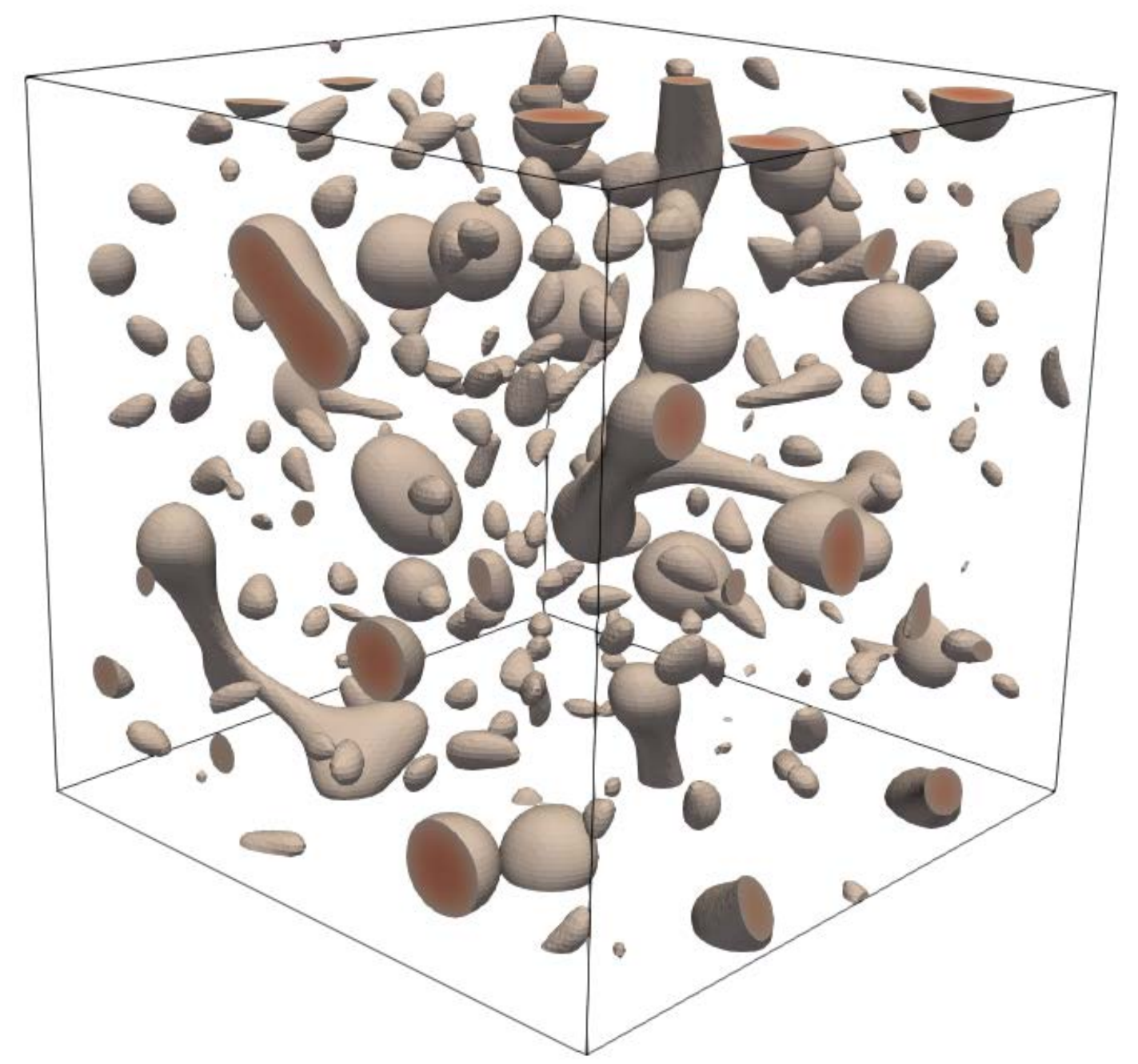}
d)\includegraphics[width=0.33\textwidth]{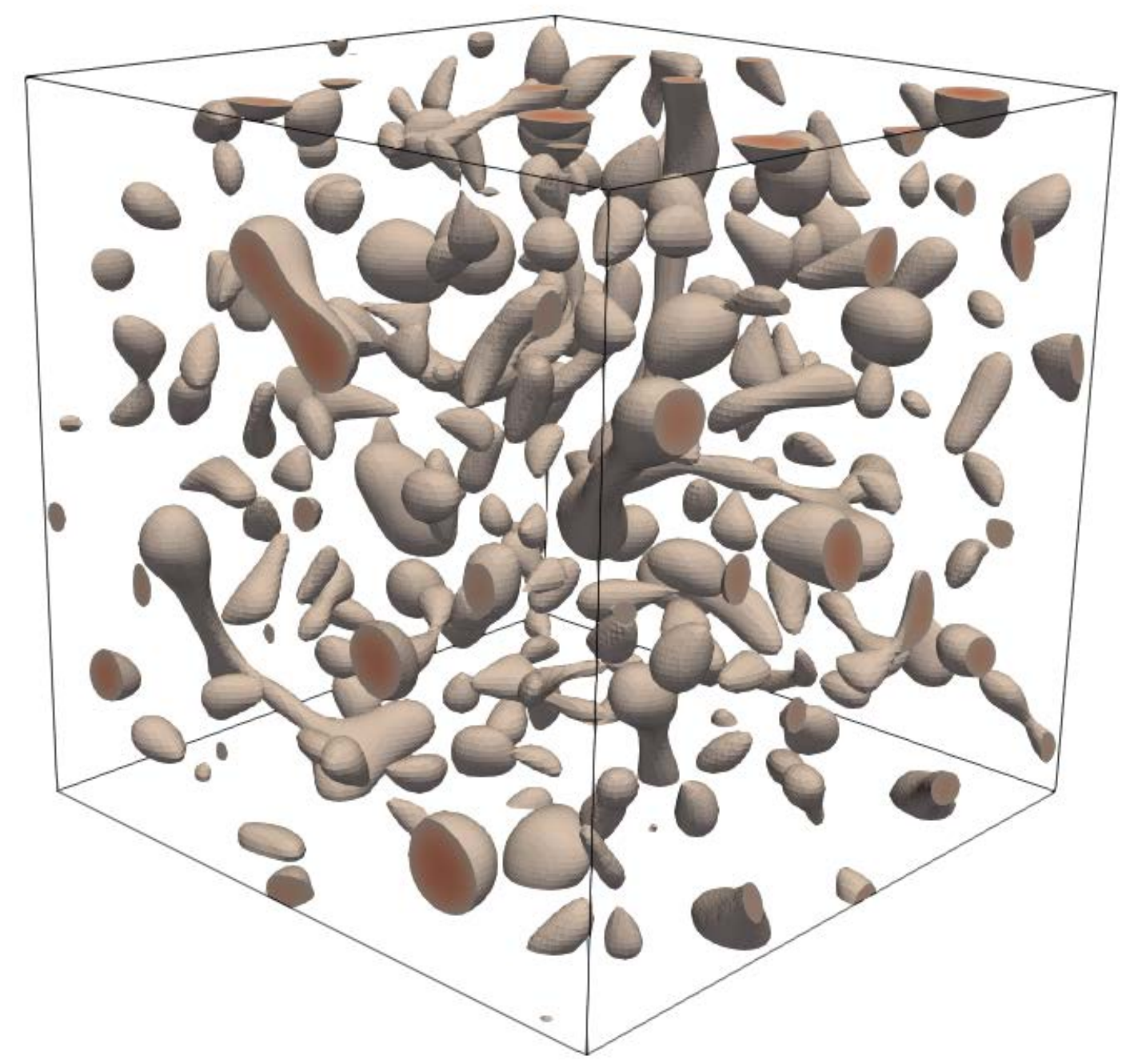}
\caption{(Colour online) Snapshots of the rearrangement of the SP particles during a  sustained irradiation at dose: a) 1~dpa; b) 2~dpa; c)  3~dpa; d) at  5~dpa.}\label{irr1}
\end{figure}

\begin{figure}
\centering
a)\includegraphics[width=0.40\textwidth]{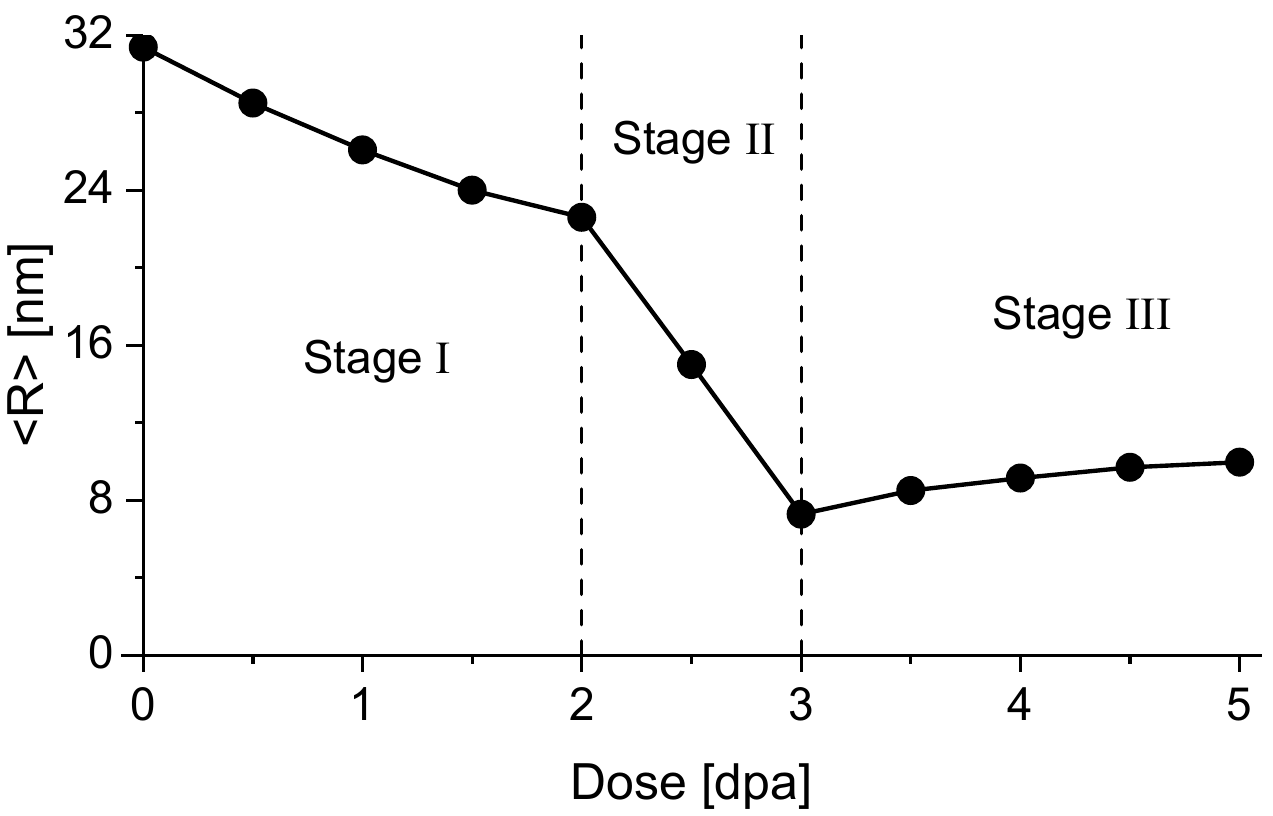}
b)\includegraphics[width=0.40\textwidth]{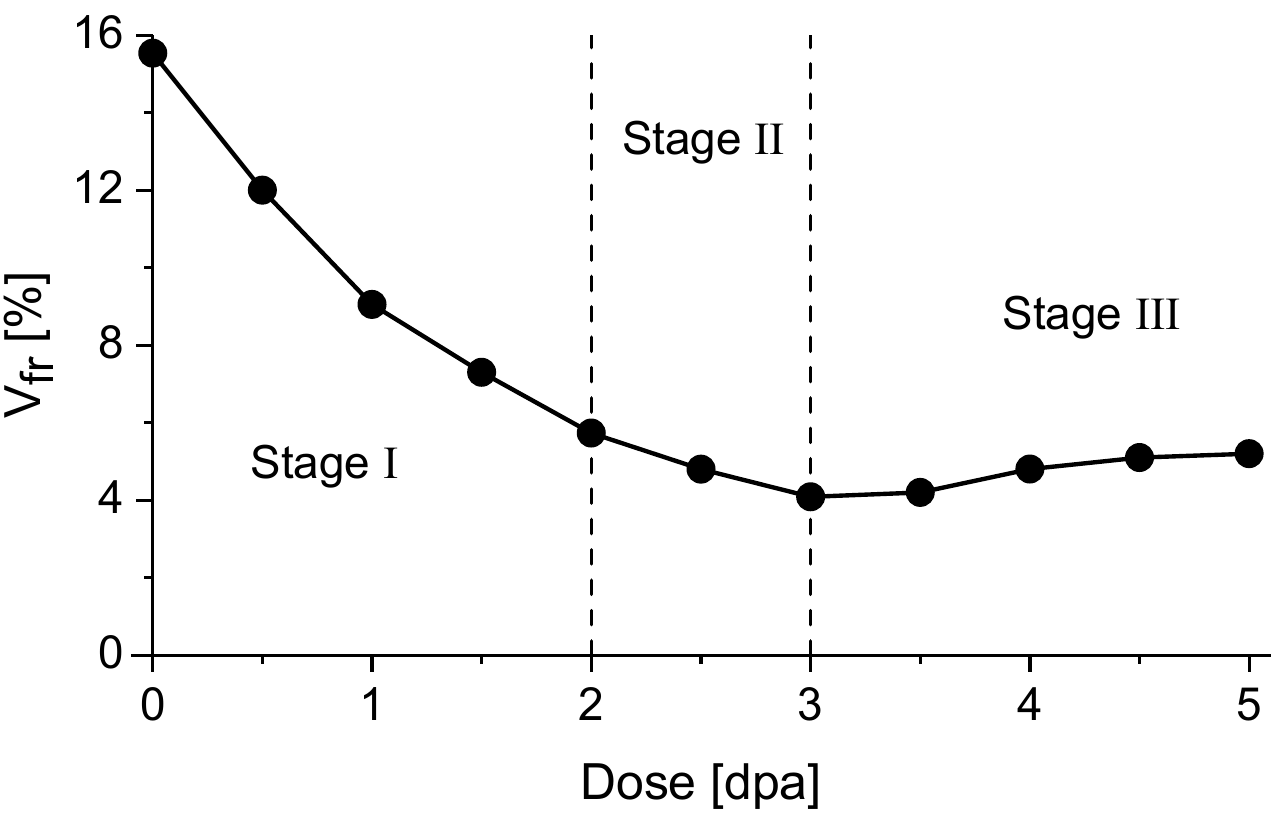}
\caption{Dose dependence of a) the mean size of SP domains; b) the volume fraction of SP.}\label{irr2}
\end{figure}

\begin{figure}
\centering
a)\includegraphics[width=0.3\textwidth]{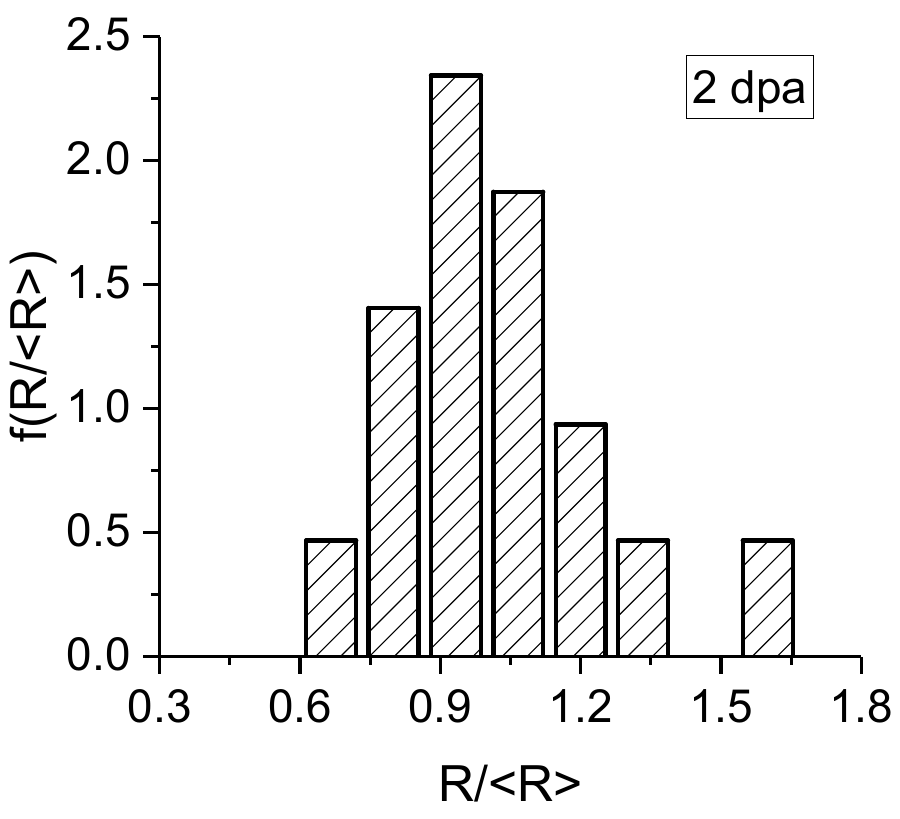}
b)\includegraphics[width=0.3\textwidth]{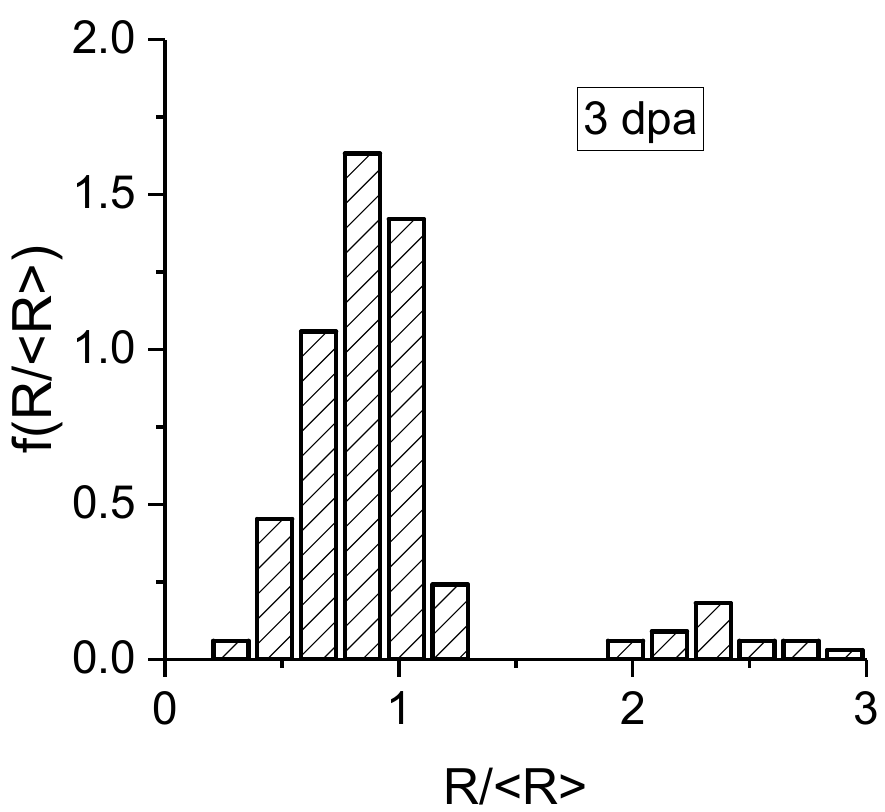}
c)\includegraphics[width=0.3\textwidth]{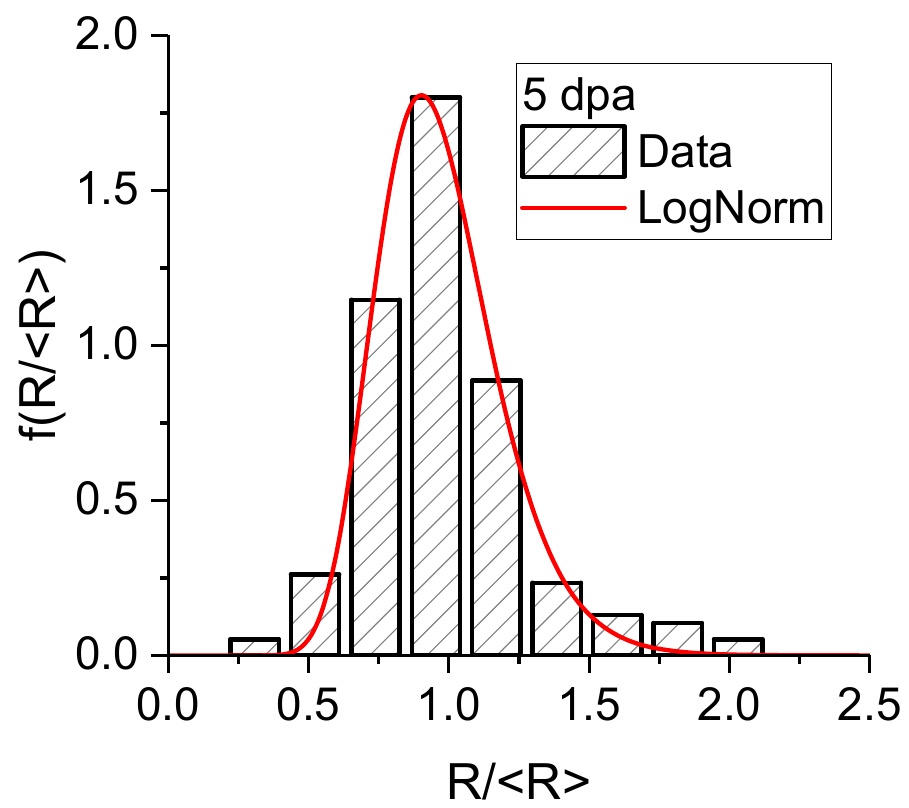}
\caption{(Colour online) Distribution of SP domains over sizes at different doses: a) 2~dpa; b) 3~dpa; c)~5~dpa.}\label{irr3}
\end{figure}

Typical scenario of microstructure transformations caused by formation of structural disorder is shown in figure~\ref{irr1}. Dependencies of the mean size (the radius of the sphere of the same area) $\langle R\rangle$ and the volume fraction of precipitates $V_{fr}$, calculated as a ratio between the area of all precipitates and the volume of the system, are shown in figure~\ref{irr2}a,b, respectively.  Distributions of SP precipitates over sizes at different doses are shown in figure~\ref{irr3}. It follows that with the irradiation dose accumulation, the native precipitates of tin-enriched SP dissolve (see snapshots a and b in figure~\ref{irr1} at 1~dpa and 2~dpa). At these doses, the mean size of precipitates and their volume fraction decrease (see figure~\ref{irr2} at doses up to 2~dpa corresponding to the Stage I).  This means that the concentration of homogeneously distributed tin in a bulk increases.  The corresponding distribution in figure~\ref{irr3}a has a uni-modal form with the most probable size in the vicinity of the mean size. 
Further irradiation results in a crucial change of the SP microstructure. Whilst the large domains of native SP continue to decrease in size, the new  irradiation-induced precipitates (IIPs) of SP particles of small size emerge (see snapshot in figure~\ref{irr1}c).  These effects result in a crucial decrease in the mean linear size of SP precipitates (see Stage II in figure~\ref{irr2}a).  At the same time, the volume fraction continues to decrease  smoothly (see figure~\ref{irr2}b). The microstructure  of SP precipitates at dose 3~dpa is characterized by the bi-modal distribution shown in figure~\ref{irr3}b. At elevated doses (over 3~dpa), the IIPs grow in size (see the snapshot in figure~\ref{irr1}d): both the mean linear size of SP particles and the volume fraction increase with the irradiation dose (see Stage III in figure~\ref{irr2}). Here, all SP particles tend to have similar sizes as it follows from their distribution over sizes in figure~\ref{irr3}c which fits well by the log-normal distribution.  

\subsubsection{Segregation of nonequilibrium vacancies}

\begin{figure}[h]
	\centering\includegraphics[scale=0.35]{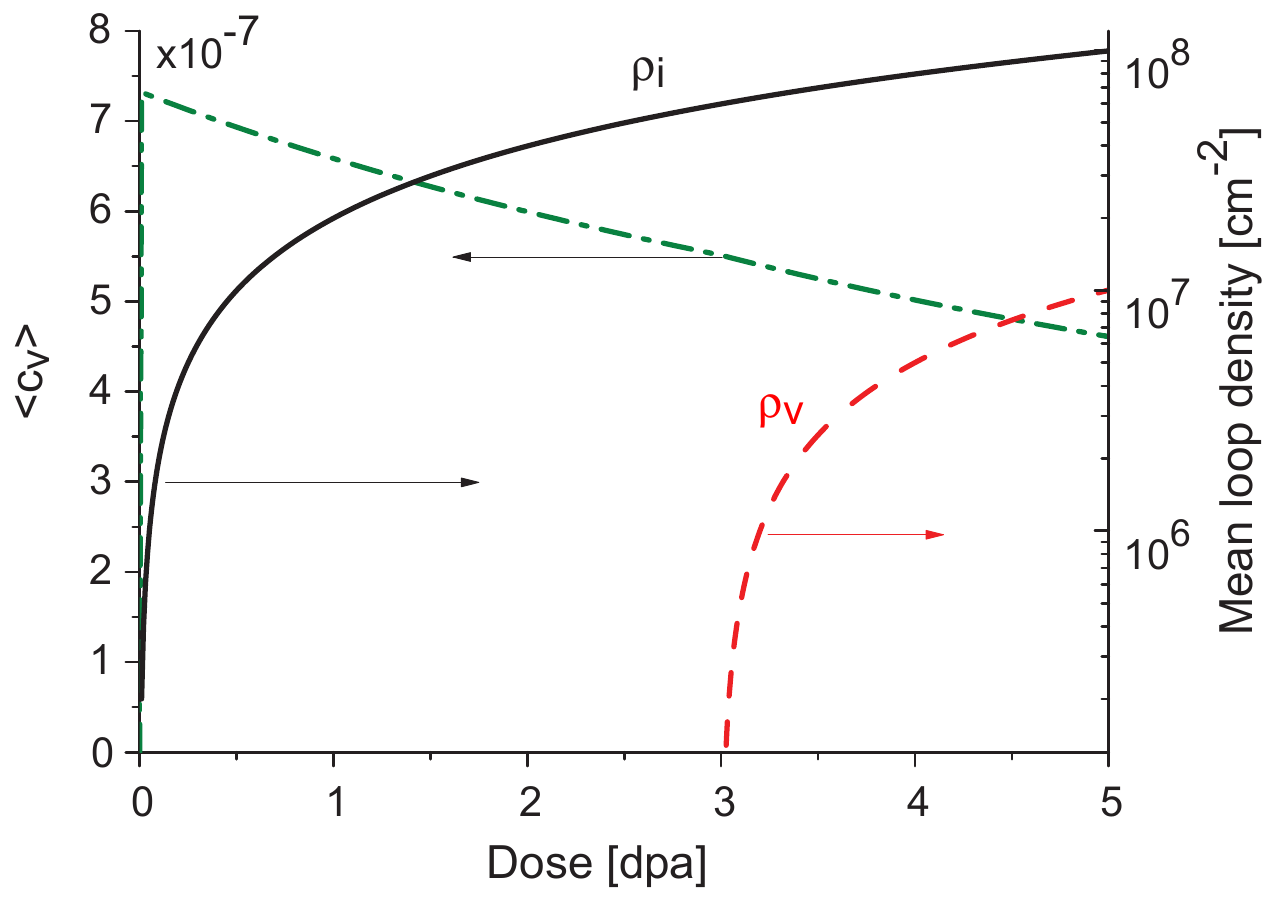}
	\caption{(Colour online) Dose dependencies of the mean concentration of nonequilibrium vacancies and mean 
		loop density, produced during irradiation.}\label{irr4}
\end{figure}

First, let us consider the dose dependencies of the mean concentration of nonequilibrium vacancies and 
loop density, produced during irradiation, shown in figure~\ref{irr4}.
From the obtained results it follows that the mean concentration of vacancies in the 
alloy (dash-dot curve in figure~\ref{irr4}) rapidly increases from the equilibrium 
value $10^{-17}$ towards $7.3\cdot10^{-7}$ at initial stages of the alloy irradiation 
due to their generation in cascades of atomic displacements. The interstitial loop starts 
to grow at small doses (solid curve in figure~\ref{irr2}); vacancy loops start 
to grow starting from the dose 3~dpa as it is shown by a dash curve. Their growth leads to a decrease in the concentration of vacancies. The obtained results for the mean vacancy concentration and 
loop density correspond to the data obtained in the framework of reaction rate theory~\cite{rrtloops} and experimental observations \cite{54loops,71loops,72loops}.

\begin{figure}[t]
\centering
a)\includegraphics[width=0.3\textwidth]{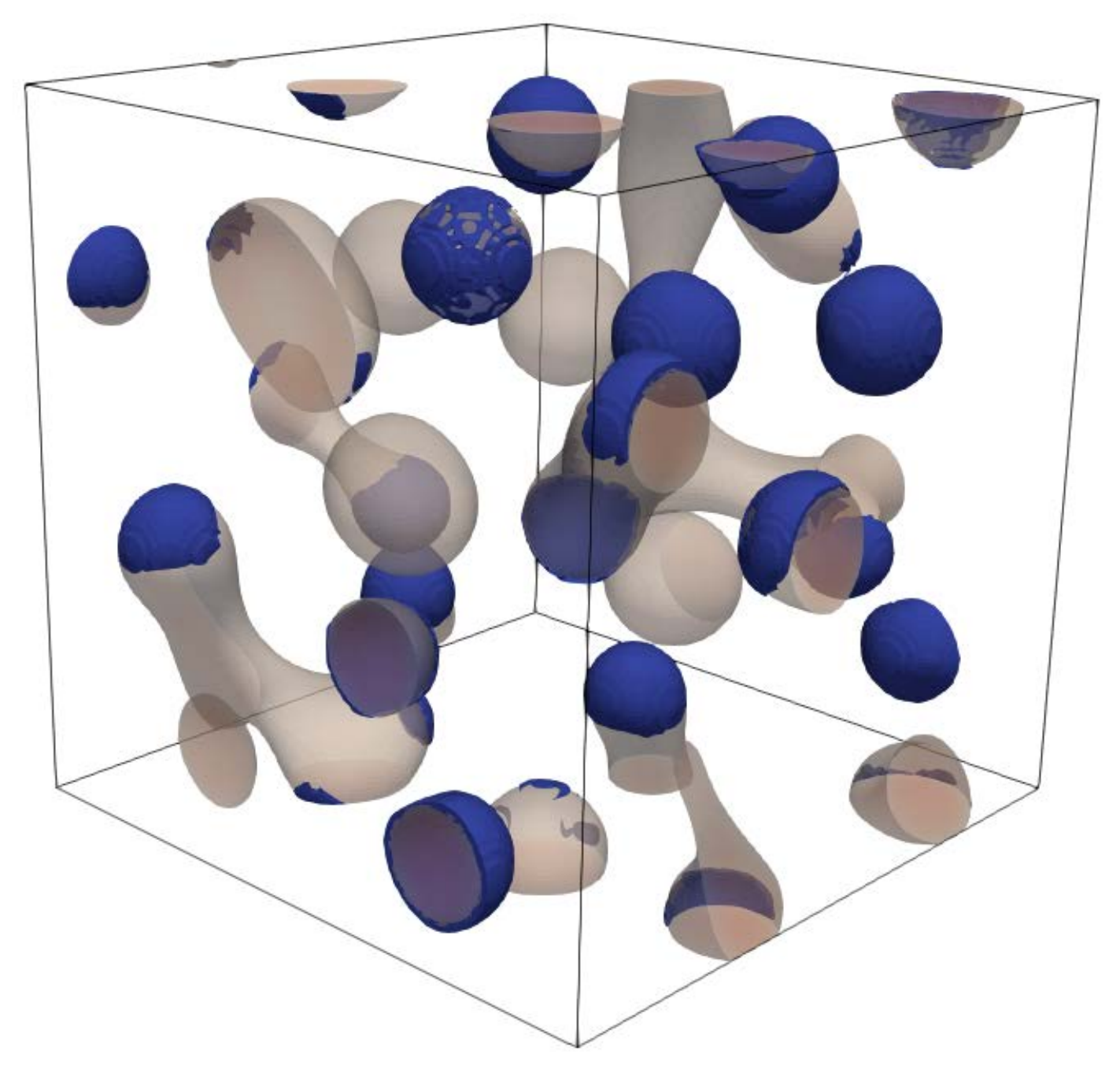}
b)\includegraphics[width=0.3\textwidth]{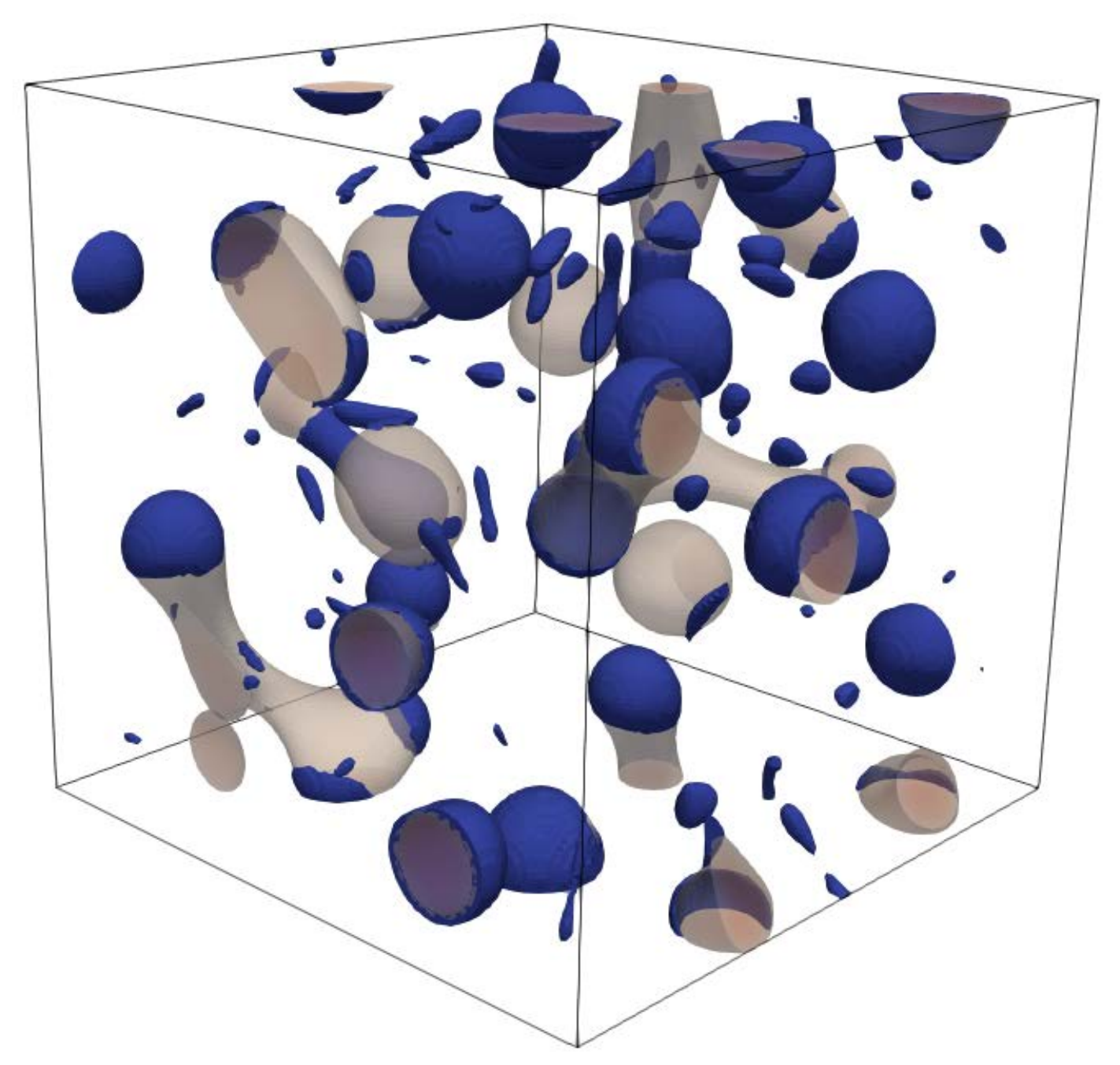}
c)\includegraphics[width=0.3\textwidth]{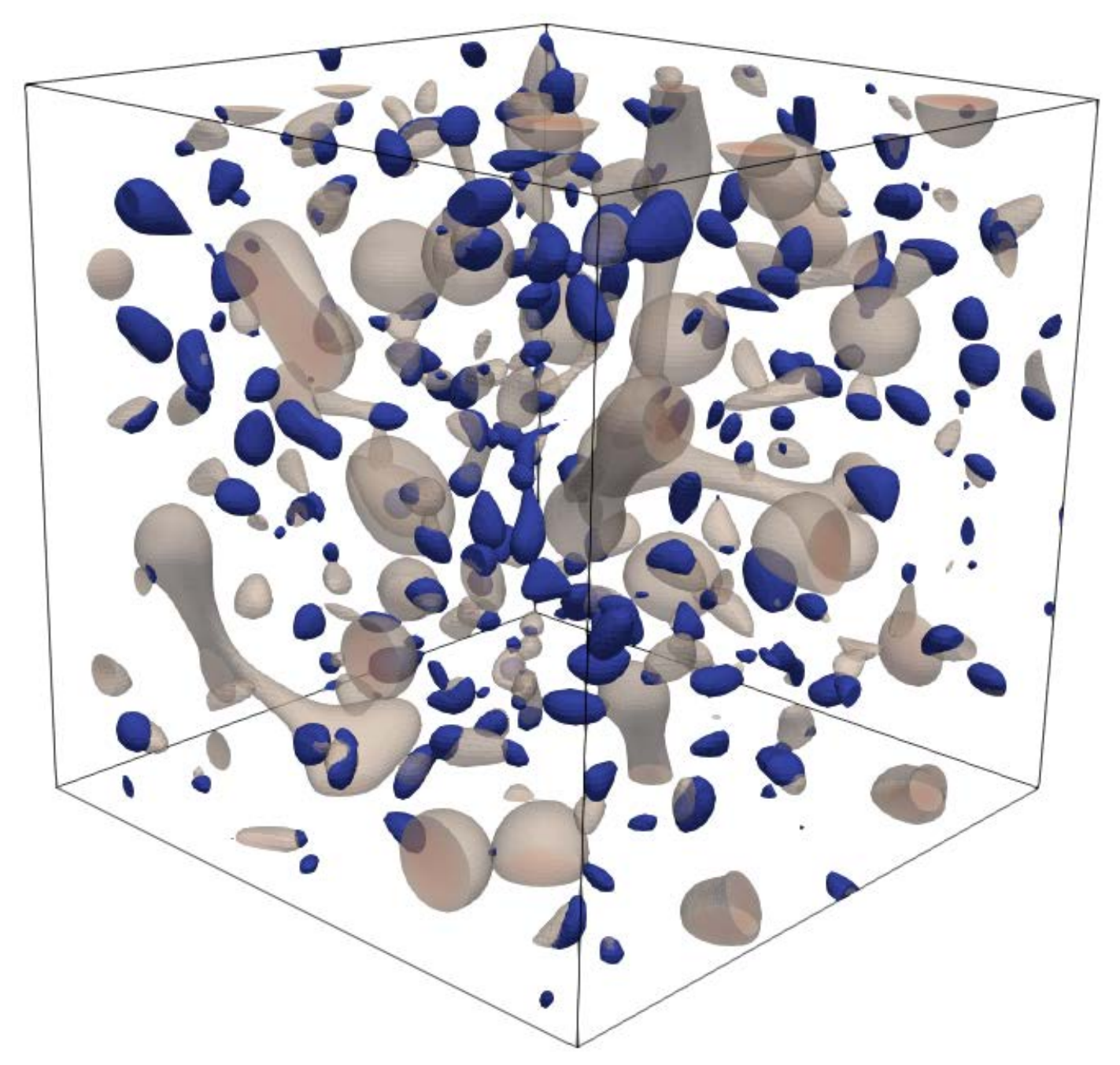}
\caption{(Colour online) Snapshots of the rearrangement of the nonequilibrium vacancies with the concentration $c_v({\bf r})>c_v^{th}$  during sustained irradiation at dose: a) 1~dpa; b) 2~dpa; c)  3~dpa.}\label{irr5}
\end{figure}

Next, let us discuss the peculiarities of spatial distribution of nonequilibrium vacancies in the studied Zr--Sn alloy under the irradiation action. It is known that  
in the irradiated systems, nonequilibrium vacancies produced in cascades of atomic displacements can form small vacancy clusters \cite{kiritani1990recoil,evans1971observations,kharchenko2013simulation,kharchenko2014modeling} and voids \cite{evans1971observations,cmph2018voids,sass1973diffraction,johnson1983bubble}.
Moreover, vacancies can be trapped by Sn atoms, meaning that the local vacancy concentration is quite related to Sn rearrangement in a bulk. In order to visualize the spatial distribution of nonequilibrium vacancy field with elevated concentration $c_v({\bf r})$, next we use the threshold $c_v^{th}=c_v^{\rm (min)}+0.85\big[c_v^{\rm(max)}-c_v^{\rm(min)}\big]$ dependent on the accumulated dose.
The distinguished vacancy enriched domains with the $c_v({\bf r})>c_v^{th}$ together with the SP domains (shown with small opacity) at different accumulated doses are shown in figure~\ref{irr5} by blue color. It follows that at small doses (see figure~\ref{irr5}a at 1~dpa) the nonequilibrium vacancies are located mostly on the phase interface of SP particles with large curvature, and the spatial distribution of vacancies does not crucially differ  from the one of unirradiated alloy (see the snapshot in figure~\ref{prep2}b). It means that at small doses, the attractive Sn-vacancy interaction plays a major role in spatial rearrangement of nonequilibrium vacancies. With the dose accumulation, one part of nonequilibrium vacancies remains on the phase interfaces and another part starts to organize into small vacancy clusters in a bulk  (see snapshot in figure~\ref{irr5}b at dose 2~dpa).   These vacancy clusters absorb tin from the matrix due to attractive interaction between Sn atoms and vacancies \cite{WU2021101765}. These effects lead to precipitation of small particles of tin enriched SP as it is seen from figure~\ref{irr1}c.  
With further irradiation, the  number of small vacancy clusters increases and their mean size grows with dose (see snapshots in figure~\ref{irr5}c,d). 

\begin{figure}[h]
\centering
a)\includegraphics[width=0.35\textwidth]{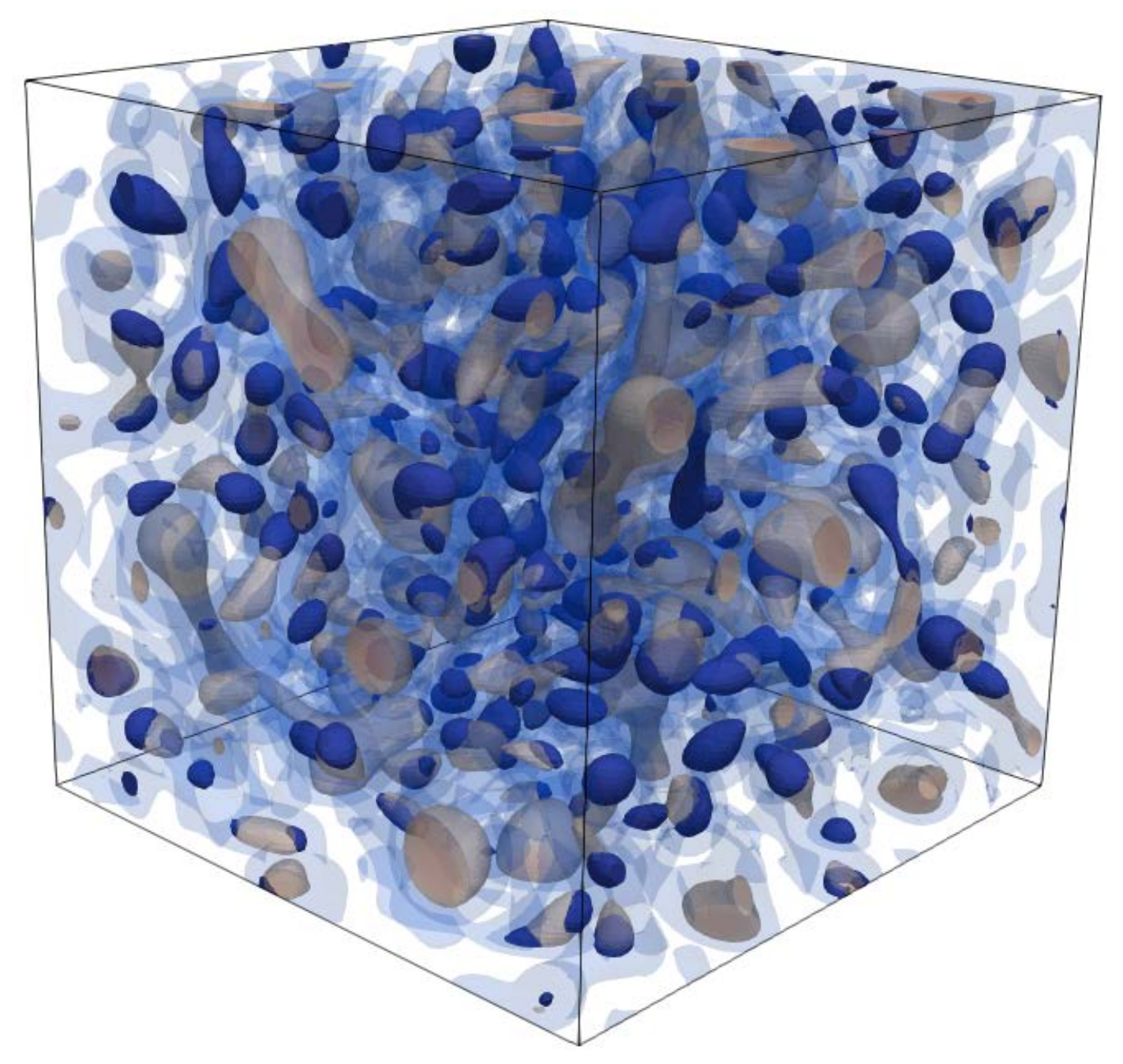}
b)\includegraphics[width=0.4\textwidth]{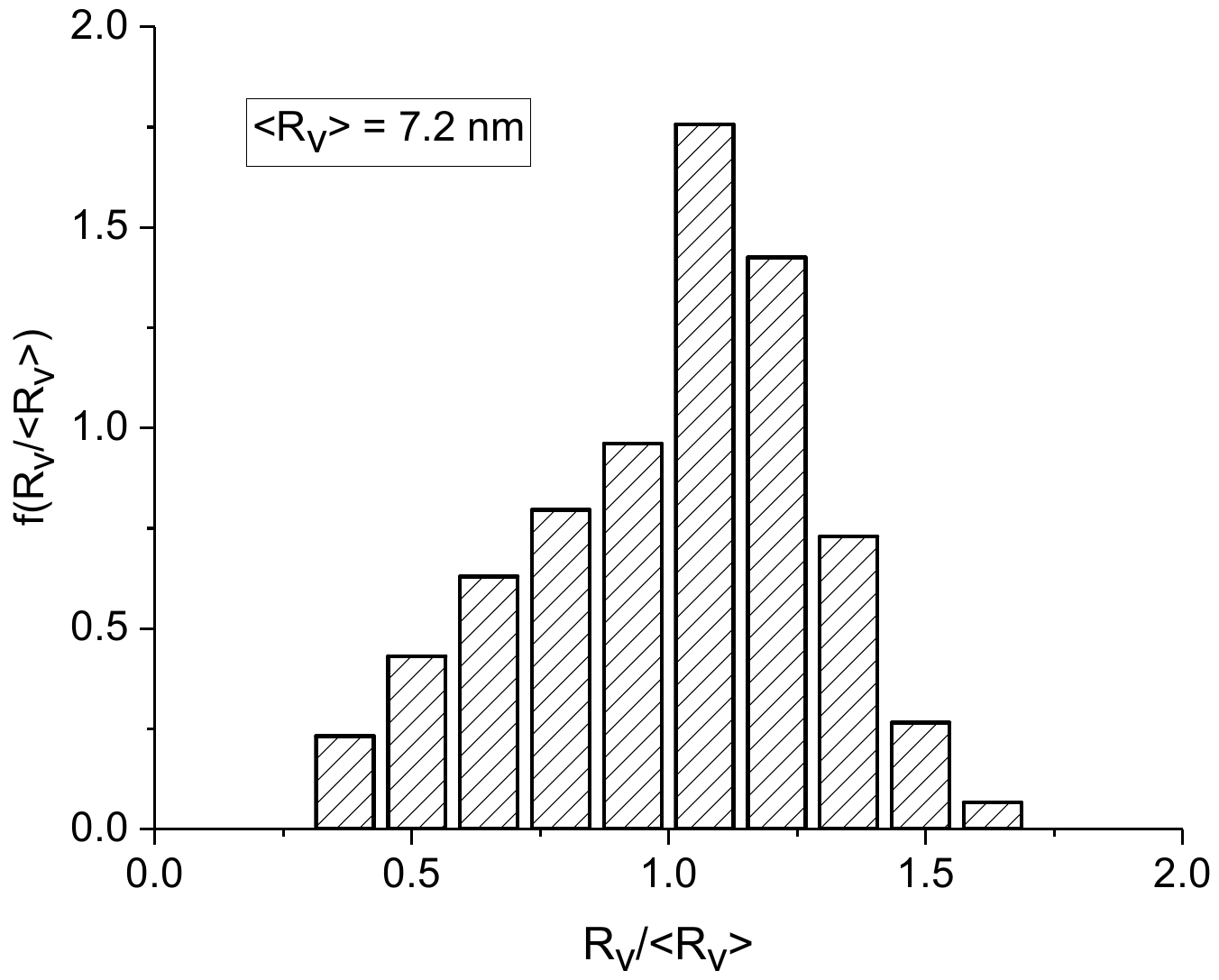}
\caption{(Colour online) Snapshot of the rearrangement of the nonequilibrium vacancies with $c_v({\bf r})>c_v^{th}$ (blue color) and with $c_v({\bf r})>\langle c_v\rangle$ (light blue color) (a) and distribution of vacancy clusters with $c_v({\bf r})>c_v^{th}$ over sizes  at dose  5~dpa (b).}\label{irr6}
\end{figure}

The vacancy clusters in figure~\ref{irr5} are characterized by the elevated concentration of vacancies $c_v({\bf r})>c_v^{th}$.  In figure~\ref{irr6}a we show the spatial distribution of vacancy concentration field with 
$c_v({\bf r})>\langle c_v\rangle$ in blue color with small opacity with the distribution of tin concentration at dose 5~dpa. It follows that a part of vacancies is distributed at phase interfaces of SP particles, whilst the vacancy clusters with elevated concentration are mostly located close to SP particles. We have calculated the distribution of the vacancy cluster with the threshold   $c_v({\bf r})>c_v^{th}$ over sizes, shown in figure~\ref{irr6}b. The mean linear size of these clusters $\langle R_v\rangle$ is approximately 7.2~nm, and the distribution is of a symmetric form around the mean size. 

\section{Conclusions}

We have studied the dynamics of microstructural transformation  and statistical properties of SP particles in Zr--10\%Sn alloy subjected to a sustained irradiation in the framework of numerical simulations in a three-dimensional system. 

It is found that in the unirradiated microstructure, obtained during thermal treatment of the corresponding solid solution at a fixed temperature $T=550$~K, the volume fraction of tin-enriched SP particles with the linear size 32~nm is around 16\% and their size distribution is close to Lifshitz--Slyozov--Wagner type. The concentration of tin in SP corresponds to A15 phase with stable configurations Zr$_5$Sn$_3$ and Zr$_4$Sn, which correspond to the known experimental results \cite{A15,item5,item6} and theoretical predictions \cite{baykov2006structural}. The obtained microstructure of the annealed alloy qualitatively relates to the experimental observation \cite{A15}, and the volume fraction of A15 phase relates quantitatively with the one calculated for the experimentally presented microstructure \cite{A15}. 

It is found that with the irradiation dose accumulation, three different stages of the microstructure transformations of atomic subsystem are realized. At small doses,  dissolution of native particles with accumulation of tin in a bulk results in a decrease in the mean size of SP particles and their volume fraction. At intermediate doses,  dissolution of native precipitates and irradiation induced precipitation of new particles of SP provides a sharp decrease in the mean size of precipitates and bimodal distribution of SP particles over sizes. At elevated doses, irradiation sustains the growth of tin-enriched domains attaining the mean size up to 12~nm and their distribution over sizes is close to log-normal. It is shown that the irradiation induced precipitation of new tin-enriched domains is caused by clusterization of nonequilibrium vacancies produced in cascades in a bulk due to attractive tin-vacancy interaction.    

\section*{Acknowledgements}

The work was supported by the Ministry
of Education and Science of Ukraine (grant
No. 0124U000551).

\bibliographystyle{cmpj}
\bibliography{cmph_Zr-Sn_biblio}

\newpage
\ukrainianpart

\title{Фазово-польове моделювання мікроструктурних перетворень у сплаві Zr--Sn під час опромінення}

\author{В. О. Харченко\refaddr{label1,label2}, Д. О. Харченко\refaddr{label1}, А. В. Дворниченко\refaddr{label2}}

\addresses{
\addr{label1} Інститут прикладної фізики НАН України, Суми, 40000, вул. Петропавлівська, 58
\addr{label2} Сумський державний університет, Суми, 40007, вул. Харківська, 116
}

\makeukrtitle

\begin{abstract}
\tolerance=3000%
У цій роботі досліджується  просторова еволюція концентрацій олова у сплаві Zr--Sn та вакансій як на стадії термічної обробки твердого розчину, так і під час його опромінення в реакторних умовах. Для цього застосовано узагальнений підхід фазово-польового моделювання, що поєднує теорію швидкостей реакцій і метод CALPHAD.
Дослідження динаміки мікроструктурних змін проведено з використанням числового моделювання у тривимірному просторі.  Розглянуто процес формування вторинної фази на етапі термічної обробки, а також досліджено її стійкість за умов нейтронного опромінення. Окрему увагу приділено аналізу впливу дози опромінення на статистичні характеристики преципітатів вторинної фази.
\keywords сплави Zr--Sn, фазово-польове моделювання, фазове розшарування, опромінення, статистичні властивості
\end{abstract}

\end{document}